\documentclass[sn-mathphys-num]{sn-jnl}

\usepackage{graphicx}%
\usepackage{multirow}%
\usepackage{amsmath,amssymb,amsfonts}%
\usepackage{amsthm}%
\usepackage{mathrsfs}%
\usepackage[title]{appendix}%
\usepackage{xcolor}%
\usepackage{textcomp}%
\usepackage{manyfoot}%
\usepackage{booktabs}%
\usepackage{algorithm}%
\usepackage{algorithmicx}%
\usepackage{algpseudocode}%
\usepackage{listings}%

\usepackage{comment}
\DeclareMathOperator{\sech}{sech}

\usepackage{xcolor}

\usepackage{booktabs}

\usepackage{lipsum}

\theoremstyle{thmstyleone}%
\theoremstyle{thmstyletwo}%

\theoremstyle{thmstylethree}%

\begin{document}

\title[Article Title]{Phase Transitions in single species Ising Models with Non-Reciprocal couplings}


\author*[1,2]{\fnm{Adrià} \sur{Garcés}}\email{adria.garces@ub.edu}

\author[1,2]{\fnm{Demian} \sur{Levis}}\email{levis@ub.edu}

\affil[1]{\textit{Computing and Understanding Collective Action (CUCA) Lab, Condensed Matter Physics Department, Universitat de Barcelona, Mart\'i i Franqu\`es 1, E08028 Barcelona, Spain}}

\affil[2]{\textit{University of Barcelona Institute of Complex Systems (UBICS), Mart\'i i Franqu\`es 1, E08028 Barcelona, Spain}}

\abstract{We present a general framework for incorporating non-reciprocal interactions into the Ising model with Glauber dynamics, without requiring multiple species. We then focus on a model with vision--cone type interactions. We solve it in a fully connected network (mean-field) and  perform extensive numerical simulations of the model in the square lattice.    
We find that the breakdown of the spin-flip symmetry introduced by non-reciprocity  induces a discontinuous phase transition on top of the usual continuous one, that eventually occurs at higher critical temperatures. 
Combining a static and dynamic scaling analysis, we measure the critical exponents associated to the continuous symmetry breaking transition, and find them to be identical  to  the ones of the Ising model in two-dimensions (2D),  with the exception of the exponent  $\beta$ associated to the order parameter. The latter appears to increase as the non-reciprocity of the coupling increases, suggesting that, within our numerical precision, the model does not belong to the 2D Ising model universality class.  The coarsening process is anisotropic, but still follows the usual dynamic scaling with a dynamic exponent  compatible with the standard  value with non-conserved order parameter dynamics. }




\maketitle
\newpage
\tableofcontents

\newpage

\section{Introduction}\label{sec1}

The description of a  large variety of non$-$equilibrium systems at coarsegrained scales, involve non$-$reciprocal interactions: such as social agents \cite{sabine_living, castellano2009statistical}, 
catalytic colloids \cite{agudo-canalejo, saha2020scalar, schmidt2019light, dinelli2023nonreciprocal}, 
multi-species ecosystems \cite{lotka-volterra, chaos_eco}, 
robotic swarms \cite{brandenbourger2019non, march2023honeybee} or 
non$-$Hermitian quantum  systems \cite{non-hermitian}. 
These systems are characterised by effective interactions between meso-scale units that appear to violate of the actio-reactio principle, 
which eventually  drive non$-$equilibrium phase transitions as well as the emergence of novel self-organisation routes and spatio-temporal patterns  \cite{traveling_Marchetti, saha2020scalar, loos_irreversibility_2020, Fruchart2021, zhang2023entropy, suchanek_23, suchanek2023irreversible, osat2023non}. 

Non-reciprocity naturally arises in muti-species systems for which the interactions are species-dependent. A typical example being a predator chasing a prey, the prey willing to escape from it   
\cite{theveneau2013chase, sabine_living}. 
From the soft matter perspective, most  work has focused on non-reciprocal coupling between  different species in a mixture \cite{ivlev2015statistical,traveling_Marchetti, saha2020scalar, Fruchart2021, suchanek2023irreversible, guislain_mean_field, Guislain_2024_spin, avni2023non}.
Another approach consists in considering non-reciprocal interactions as arising from  interactions that explicitly depend on the internal state of the elementary constituents of the system. Such feedback  is motivated by  epidemic/opinion spreading models 
and animal groups interacting with their peers within a finite vision range 
 \cite{couzin2003self, peruani2008dynamics, bastien2020model, levis2020flocking,rodriguez2022epidemic, gomez2022intermittent, seara2023, rajeev2024ising, di2024off}.

How non-reciprocal interactions influence classical phase transitions remains an open question. From the perspective of out-of-equilibrium dynamics, understanding the slow, glassy behavior of non-reciprocal systems has been a challenge since it was first posed decades ago in the context of neural networks and spin-glass theory \cite{CrisantiIsing, CrisantiSpherical, parisi1986asymmetric}. This topic has recently gained renewed interest,  particularly due to advances in active matter, biophysics and theoretical ecology research \cite{Guislain_disordered, guislain2024far, hanai2024nonreciprocal,lorenzana2024non}.
For single component systems, recent works have addressed this question in the two-dimensional (2D) XY model, 
introducing vision-cone ferromagnetic couplings   \cite{durve2018active, loos2023long, rouzaire2024non, huang2024visioncones}.  Here we investigate the impact of non-reciprocal couplings in the symmetry breaking phase transition scenario of the 2D and mean-field Ising model. To do so, we introduce a general  non-reciprocal Ising model that allows us to make a smooth  connection with the standard, reciprocal  model,  without introducing multiple  species.

Being the canonical model to study spontaneous symmetry breaking phase transitions, Ising models with non--reciprocal interactions have been  recently considered  \cite{avni2023non, seara2023, rajeev2024ising, guislain_mean_field, Guislain_2024_spin}. In \cite{avni2023non}  a two-species model is analyzed, where each spin species interacts  antisymetrically with the opposite species spins: spins of, say, species A, tend to align to spins B (ferromagnetic coupling), while the latter tend to anti-align with spins of the opposite species  (anti-ferromagnetic coupling).   
This gives rise to a diverse map of stationary phases, including a "chase and run" dynamical phase in which the orientation of spin pairs of opposite species are persistently unstable. This oscillatory phase is shown to disappear in $d = 2$ through the formation of spiral defects,  while it remains stable in $d = 3$. A similar approach is taken in \cite{guislain_mean_field, Guislain_2024_spin} in which the two species consist on spins $\sigma_i$ and fields $h_i$:  
spins (fields) interact reciprocally with other spins (fields) in their neighbourhood, while the interaction of spins (fields) with its neighbouring fields (spins) is non-reciprocal.
Single--species models in which the non--reciprocity is introduced by breaking the isotropy of the Ising model's interaction matrix have been recently explored in \cite{rajeev2024ising, seara2023, di2024off}. Both \cite{rajeev2024ising, seara2023} report the emergence of traveling waves whose velocity and direction is controlled by the parameter breaking  reciprocity: changing the coupling asymetrically  \cite{rajeev2024ising}, and introducing  spatial anisotropy in the coupling  \cite{seara2023}. In both cases the authors introduce a bias in the coupling that does not depend on the state of the spins, thus globally advecting the magnetisation field along an imposed direction.
 The presence of  advection and its effect are studied in \cite{di2024off}.



In this work, we introduce non-reciprocity in the Ising model on general grounds, by considering exchange couplings that explicitly depend on the state of the interacting spins. As the energy of a bond is an ambiguous quantity in this context, we take a purely dynamical definition of the model, taking as a starting point the kinetic Ising model with Glauber dynamics \cite{glauber1963time}. Within such framework, we  first show how at the mean-field level,  the non-reciprocity of the interactions can break the spin inversion symmetry, leading to diverse phenomena such as metastability and hysteresis. We further show how in finite dimensions the continuum description 
 of hypercubic $d-$dimensional lattices can incorporate non--equilibrium terms into the dynamics, which do not derive from a Ginzburg--Landau functional. 

We investigate what happens when spins interact non--reciprocally through a vision--cone. 
We find a shift in the critical temperature which can be predicted using the mean--field results in the weakly non--reciprocal limit as well as the emergence of a first order phase transition. 
 Interestingly, a finite-size scaling analysis reveals  a change in the critical exponent $\beta$ while other critical exponents $\nu, \gamma$ and $z$ remain consistent with the universality class values of the Ising model in $d = 2$ with non-conserved order parameter dynamics (or model A in the continuum picture) \cite{bray2002theory}.
The change of $\beta$ with non--reciprocity is consistent with the analysis of the short-time dynamics near the critical temperature, which provides information about both static and dynamic critical exponents.  

This article is organized as follows, in Sec. \ref{sec:2} we introduce the general model that incorporates non--reciprocal state--dependent interactions between Ising spins endowed with Glauber (Markovian) dynamics. We study fully connected models and derive a general evolution equation for the magnetization of the system in the weakly non--reciprocal model and show how one can build a continuum description in finite dimensional lattices. In Sec. \ref{sec:3} we particularize the general model to incorporate vision--cone interactions. We study analytically the phase behaviour at the mean--field level, we present the continuum theories that result from considering vision--cone interactions in finite dimensions  and we present the simulation results of the model in the square lattice. In Sec. \ref{sec:4} we study the phase ordering kinetics of the model.  Sec. \ref{sec:5} closes the article with a discussion of the main findings and open questions.

\section{Non-Reciprocal Ising Models with state-dependent coupling}\label{sec:2}

\subsection{Lattice Models}
\begin{center}
    \begin{figure}[t]
        \centering
        \includegraphics[width=0.9\textwidth]{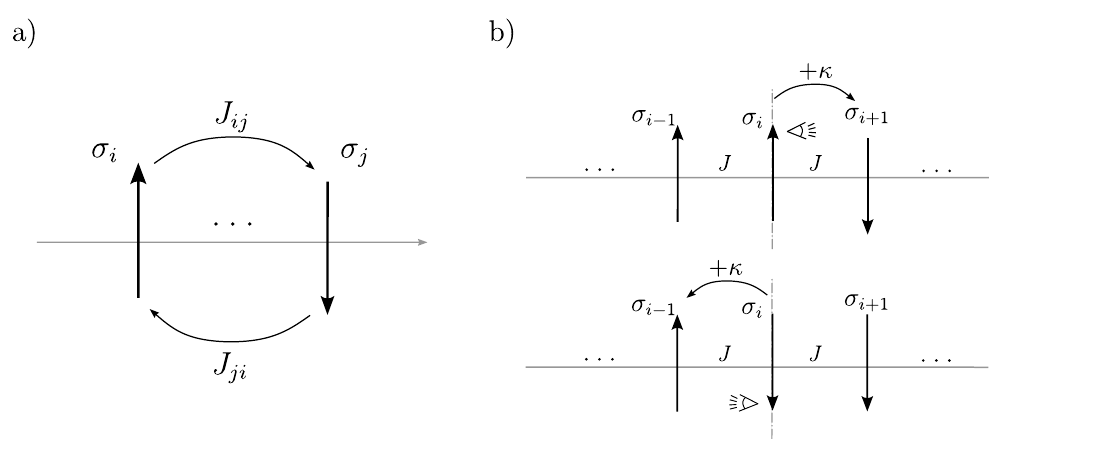}
        \vspace{-0.3cm}
       \caption{(a) Sketch of a single-species non-reciprocal interaction. (b) Schematic drawing of non$-$reciprocally interacting spins in a $d=1$ chain inspired by vision cones. Spins interact with two different couplings, a reciprocal one $J$ which represents nothing but the coupling constant of the Ising model on top of the one due to their vision field, $\kappa$.} 
        \label{fig:sketch}
    \end{figure}
\end{center}

We consider $N$ Ising spin variables $\sigma_i(t)=\pm 1$ at time $t$, sitting on the nodes $i=1,\dots, N$ of a 
$d$-dimensional  lattice. The model is then defined by its stochastic dynamics, obeying the following master equation 
\begin{equation}\label{eq:master_eq}
    \frac{d p(\boldsymbol{\sigma};t)}{d t} = \sum_{i=1}^N \left[\omega(-\sigma_i)p(\boldsymbol{\sigma}^i; t) - \omega(\sigma_i) p(\boldsymbol{\sigma}; t)\right],
\end{equation}
where $p(\boldsymbol{\sigma}; t)$ is the $N$-body probability of finding the system at a given configuration $\boldsymbol{\sigma} := (\sigma_1,\dots,\sigma_i,\dots,\sigma_N)$ at time $t$.
We denote  $\boldsymbol{\sigma}^i := (\sigma_1,\dots,-\sigma_i,\dots,\sigma_N)$ the configuration resulting from flipping $\sigma_i$  from state $\boldsymbol{\sigma}$, and $\omega(\sigma_i)$ the transition rate associated to such a single spin-flip. 
We introduce non-reciprocal couplings  on general grounds by specifying the following Glauber transition rates   \cite{glauber1963time},
\begin{equation}\label{eq:glauber_rates}
    \omega(\sigma_i) = \frac{1}{2}[1-\sigma_i \tanh (\beta h_i)],
\end{equation}
where $\beta = 1/k_B T$, and  
\begin{equation}\label{eq:local_field}
    h_i =h_i^{\text{ext}} + \sum_{j=1}^{N}J_{ij}\sigma_j.
\end{equation}
While $h_i^{\text{ext}}$  stands for an external conjugated field acting on  $\sigma_i$, the $J_{ij}$ matrix defines the interaction between spins and it is where non-reciprocity will come into play. 
In the standard Ising model, such transition rates come from the requirement of fulfilling detailed balance, ensuring convergence towards the equilibrium distribution $p_0(\boldsymbol{\sigma})\propto e^{-\beta H(\boldsymbol{\sigma})}$ with $H(\boldsymbol{\sigma})=-\sum_{i,j}J_{ij}\sigma_i\sigma_j -\sum_ih_i^{\text{ext}}\sigma_i$.  
Here we consider a rather general extension where $J_{ij}$ can be split  into a symmetric and an asymmetric part, namely $J_{ij}=J^s_{ij}+J^a_{ij}$, with $J^s_{ij}=J^s_{ji}$ but $J^a_{ij}\neq J^a_{ji}$, thus breaking the \textit{actio-reactio} principle (see sketch Fig. \ref{fig:sketch}).  Such couplings are naturally  non-reciprocal if they depend on the dynamic variables themselves. 
We then  set $J^a_{ij}=\kappa \rho_{ij}(\boldsymbol{\sigma})$, where $\rho_{ij}=0,\pm 1$, $\forall (i,j)$, and $\kappa$ defines the \emph{non-reciprocal parameter}  quantifying the departure from the standard, reciprocal case. Contrary to other models, the present approach does not rely on multiple species  and it is somehow closer in spirit to vision-cone type interactions (although not restricted to it, as we'll discuss after). That said,  the present framework is general enough to also allow for  multispecies models (see Appendix \ref{sec:app_a}).
The local field acting on a given spin can now be written as 
\begin{equation}\label{eq:non_reciprocal_field}
    h_i =  h_i^{\text{ext}}+ \sum_{j=1}^{N}J_{ij}^s \sigma_j + \kappa \sum_{j=1}^{N}\rho_{ij}(\boldsymbol{\sigma})\sigma_j, 
\end{equation}
and together with the Glauber rates in Eq. (\ref{eq:glauber_rates}), it sets our general framework to systematically define non-reciprocal lattice models. 
From now on, we restrict ourselves to $ h_i^{\text{ext}}=0$.

\subsection{Fully Connected \& Weakly Non$-$Reciprocal limit}

To gain analytical insight on the impact of non-reciprocity on the Ising model, we start  by defining the model on a fully connected (FC) network. This will allow us to settle a mean-field picture  that should be helpful as a reference scenario to then move to critical phenomena in finite dimensions. 

As usual, the symmetric coupling of the FC Ising model is set to  $J_{ij}^s = J(1-\delta_{ij})/N$, with $J>0$ and $\delta_{ij}$ being the identity matrix. The asymmetric part $\rho_{ij}$   will be left arbitrary for the moment but fulfilling $\rho_{ii} = 0$. 
The field $h_i$ is now global since all the spins interact with each other. It is useful to disentangle the field attributed to the reciprocal coupling, $h^s_i = JN^{-1}\sum_{j \neq i}\sigma_j$, from the non-symmetric contribution. 

We now consider only weak departures from the reciprocal reference Ising model and perform a perturbation expansion in $\kappa$. This will allow us to linearise the transition rates and derive analytical expressions for some thermodynamic observables of interest, such as the order parameter.  
The limit  $|\kappa| \ll J$ can be taken by expanding the hyperbolic tangent term  in Eq. (\ref{eq:glauber_rates}) to first order in $\kappa$, 
\begin{equation}
    \omega(\sigma_i)  = \omega_s(\sigma_i) +\kappa \delta\omega_i + \mathcal{O}(\kappa^2),
\end{equation}
where $\omega_s(\sigma_i) = \frac{1}{2}[1-\sigma_i \tanh \beta h_s^i]$ is the spin flip rate of the reciprocal FC Ising model, and where
\begin{equation}
    \delta\omega_i = -\frac{\beta \sigma_i}{2}\left(\sum_{j=1}^{N}\rho_{ij}\sigma_j\right) \sech^2 \beta h_s^i
\end{equation}
is the first order correction that non-reciprocity brings to the Glauber rates. 
The evolution of the expected value $\langle \sigma_i \rangle$ can then be computed and gives (see Appendix \ref{sec:me-sfd}), 
\begin{equation}\label{eq:dyn_si}
    \frac{d\langle \sigma_i \rangle}{dt} = -2 \langle \sigma_{i}\omega(\sigma_i)\rangle.
\end{equation}
If one now assumes  $\langle \sigma_i\sigma_j\rangle = \langle \sigma_i\rangle\langle\sigma_j\rangle$ and defines the magnetization $m =  N^{-1} \sum_{i}\sigma_i $ - such that $h_s^i \approx Jm$  -  one can write, 
\begin{equation}\label{eq:evol_m}
    \frac{dm}{dt} = \varphi_0(m,K) + \kappa \tilde{\varphi}(m,K), 
\end{equation}
where $K=\beta \, J$ and
\begin{align}
    \varphi_0(m,K) &= - m + \tanh Km \\
    \tilde{\varphi}(m,K) &=  \frac{\beta}{N}\sech^2 Km \sum_{i=1}^{N}\sum_{j\neq i}\langle \rho_{ij}(\boldsymbol{\sigma}) \rangle \langle \sigma_j \rangle\, .
\end{align}
We recognize in $\varphi_0$ the self-consistent equation for the magnetisation of the mean-field Ising model. The other term, $\tilde{\varphi}$, is assumed to be a function of $m$ and $K$ only, as it will be shown with a specific example in the following.  It is worth stressing at this point that, while $\varphi_0$ is an odd function, resulting in  symmetric solutions of the self-consistent equations under the transformation $m\to -m$, $\tilde{\varphi}$ does not have to exhibit any specific symmetry. Indeed  the symmetry of $\varphi_0$  reflects the $\mathbb{Z}_2$ symmetry of the Ising model Hamiltonian. Non-symmetric couplings might break such symmetry in an arbitrary way, that will depend on the specific choice of $\rho_{ij}$, encoding the symmetries, if any, of the non-reciprocal model.  This will obviously have qualitative consequences in the large scale physics. 

Further assuming that $\tilde{\varphi}$ is integrable,  Eq. (\ref{eq:evol_m}) can be formally rewritten as 
\begin{equation}
     \frac{dm}{dt} = - \frac{\partial F}{\partial m} , 
\end{equation}
such that the stationary points of $m$ are the maxima and minima of the function 
\begin{equation}\label{eq:F_MF}
      {F}(m,K,\kappa) =  {F}_0 + \frac{1}{2}m^2 - \frac{\ln (\cosh Km)}{K} - \kappa \int_{0}^{m} dm'\; \tilde{\varphi}(m',K). 
\end{equation}
where $F_0$ is just a temperature dependent integration constant. In the reciprocal case, $\kappa  = 0$, $F$ is nothing but the Ising model's mean-field free energy. It is also important to note how $\varphi_0(m,K) = - \varphi_0(-m,K)$ while $\tilde{\varphi}(m,K)$ is not \emph{a priori} constrained to fulfill any symmetry. This incorporates important changes even for $|\kappa| \sim 0$ since now the stationary solutions do not have to be  $\mathbb{Z}_2$-symmetric. The steady state values of $m$ are given by the solutions of the self-consistent equation
\begin{equation}\label{eq:pp_sce}
     m = \tanh Km + \kappa \tilde{\varphi}(m,K).
\end{equation}
Obviously, to explicitly solve such equation, one needs to specify the couplings $\rho_{ij}$, as will be done in Sec. \ref{sec:3}.  At this level, the mean-field treatment is general, under the before mentioned mathematical assumptions on $\tilde{\varphi}$.  In the following, we introduce the model in finite dimensional  lattices, for which an analytical treatment as the one presented above, is out of reach. 


\subsection{Lattice model in finite dimensions}
We consider now spins laying on the nodes of a $d-$dimensional hypercubic lattice ($Z^d$). Only spins sitting in nearest-neighbouring sites interact: $J_{ij}^s = J$ if $j$ and $i$ are nearest neighbours and $J_{ij}^s = 0$ otherwise, and also restrict  $\rho_{ij}$ to be different than zero only for nearest-neighbour sites. The  evolution equation of the magnetization will now incorporate spatial fluctuations, absent in the {fully connected} treatment.  Eq. (\ref{eq:dyn_si}) can be rewritten as
\begin{equation}
    \frac{d\langle \sigma_i \rangle}{d t} = - \langle \sigma_i \rangle + \langle \tanh( \beta h_i) \rangle.
\end{equation}

The key difficulty to derive a closed set of equations is the non-linearity of the Glauber rates, i.e. the hyperbolic tangent, together with the extra correlations introduced by non-reciprocal state-dependent couplings. Defining now the local magnetization $m_i (t) = \langle \sigma_i (t) \rangle$ and assuming  $\langle \sigma_i \sigma_j\rangle = \langle \sigma_i\rangle\langle\sigma_j\rangle$ --  such that $\langle \tanh \beta h_i \rangle = \tanh \beta \langle h_i \rangle$  (see Appendix \ref{sec:mf-analytic-function}) --  one finds 
\begin{equation}\label{eq:finite_d_mf_evo}
    \frac{d m_i}{d t} = -m_i + \tanh \left(\beta J \sum_{j \in \langle i \rangle} m_j + \beta \kappa \sum_{j\in\langle i \rangle}\langle\rho_{ij}(\boldsymbol{\sigma})\rangle m_j\right).
\end{equation}
Eq. (\ref{eq:finite_d_mf_evo}) couples the local magnetization at site $i\in Z^d$ to the local magnetization of the neighbouring sites $j$, for any $i = 1,\dots, N$. One can transform these $N$ coupled equations to a continuum equation by introducing a magnetization field $m(\textbf{x},t)$ defined over $\mathbb{R}^d$ instead of the lattice $Z^d$. The magnetization at a given site $i$ is $m(\textbf{x}_i,t)$ with $\textbf{x}_i = \sum_{\alpha=1}^{d}i_{\alpha}\hat{e}_{\alpha} \in \mathbb{R}^d$. Here, $\{\hat{e}_{\alpha}\}_{\alpha=1}^{d}$ is simply the cartesian base of $\mathbb{R}^d$ and $|\hat{e}_{\alpha}| = a$ is the lattice spacing. This can always be done for the reciprocal part of the interactions, for which 

\begin{equation}
    J \sum_{j\in \langle i \rangle}m_j  \rightarrow 2dJ m(\textbf{x}_i)  + a^2J \nabla^2 m(\textbf{x}_i).
\end{equation}

However, such simple continuum approximation cannot generally be carried out for the non--reciprocal part of the interactions, because of the presence of $\rho_{ij}(\boldsymbol{\sigma})$ in the sum. In order to proceed, a particular definition of $\rho_{ij}(\boldsymbol{\sigma})$ is required. As we will see in Sec. \ref{sec:finited_z2}, \ref{sec:finited_no_z2}, for an appropriate and still physically meaningful choice of couplings, a continuum approximation of the contribution of the non--reciprocal part of the interactions can be derived. Then, formally, an expansion of the hyperbolic tangent for small $m$ and up to first order in gradients, yields

\begin{equation}\label{eq:pp_m_field_pde}
    \frac{\partial m(\textbf{x},t)}{\partial t} = -\frac{\delta F[m]}{\delta m(\textbf{x},t)} + {\mathcal{L}}[m],
\end{equation}
where the term $\delta F[m]/\delta m(\textbf{x},t)$ comes from the reciprocal part (a $m^4$-action) while the term $\mathcal{L}[m]$ comes from the non--reciprocal part of the interactions and can have terms that do no preserve the $m(\textbf{x},t)\rightarrow-m(\textbf{x},t)$ symmetry and are non--linear in gradients.

\begin{figure}
    \centering
    \includegraphics[width=0.8\linewidth]{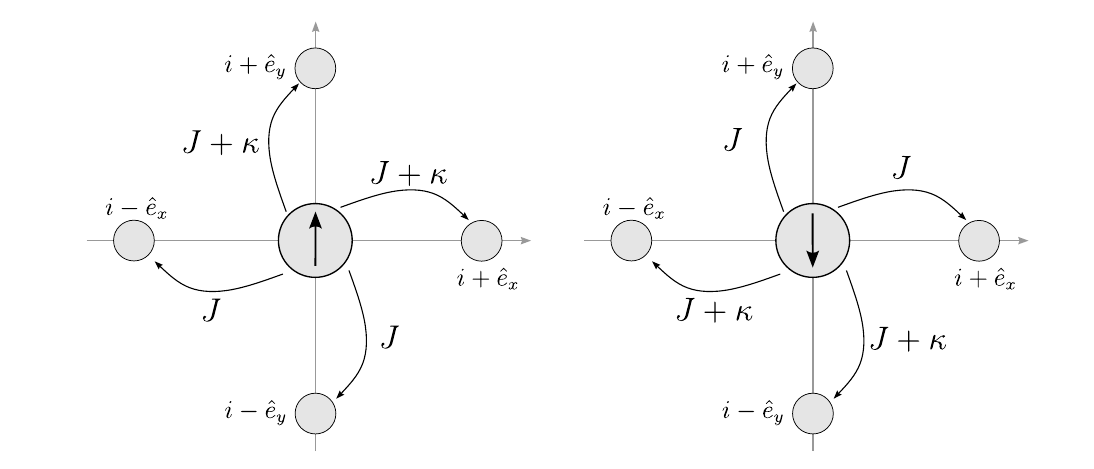}
    \caption{Illustration of the vision cone interactions in a square lattice for $f = 1/2$. Note how for $d\geq 2$ there are a number of choices depending on the fraction of neighbours $f$ to which one adds a quantity $+\kappa$ to the coupling strength $J$.}
    \label{fig:square}
\end{figure}

\section{Phase behaviour of Vision-Cone Ising Models}\label{sec:3}

\subsection{The vision-cone Ising model}

 We consider vision-cone like interaction: spins pointing in a given direction are more likely to align with spins in a restricted neighbourhood, set by the orientation of the spin itself. The coupling between a spin with its neighbourhood, thus depends on its state.
To be more concrete, let us first define the model in $1d$, as illustrated in Fig. (\ref{fig:sketch}, b). The couplings are such that spins 'up' (+1) have higher tendency to align with the spin sitting on their right than on their left. On the contrary, spins 'down' ($-1$) have higher tendency to align with spins on their left. This can be formalised by the following coupling 
\begin{equation}\label{eq:VC_FC}
\rho_{ij}=\delta_{\sigma_i,-1}\Theta(i-j) + \delta_{\sigma_i,+1} \Theta(j-i)\end{equation}
where $\Theta$ is the Heaviside step function and  $\delta$ is the Kronecker delta.

When moving to $2d$, one has to set a preferred direction of coupling for $\pm 1$ spins. On a square lattice, one has $z=4$ possible bonds to which one can eventually associate different weights to break reciprocity. 
By extension of the $1d$ case, we choose to associate a higher weight to a fraction $f$ of the bonds, depending on the state of the spin $\sigma_i$ considered.  
For instance, for $f=1/2$, we add an extra coupling $\kappa$ to top and right bonds if $\sigma_i=+1$, while such extra weight is added to bottom and left bonds if $\sigma=-1$ (see Fig. \ref{fig:square}). 
In this case, the asymmetric couplings  read
\begin{equation}\label{eq:rhoVC}
    \rho_{ij} = \delta_{\sigma_i,+}(\delta_{i+\hat{e}_x,j} + \delta_{i+\hat{e}_y,j}) +  \delta_{\sigma_i,-}(\delta_{i-\hat{e}_x,j}+\delta_{i-\hat{e}_y,j})
\end{equation}
where $\{\hat{e}_x, \hat{e}_y\}$ denote the two orthogonal lattice directions. Note that such non-reciprocal coupling can be generalized to any linear combination $\bold{v}=\lambda_x\hat{e}_x+\lambda_y\hat{e}_y$. Thus, in this notation, for the conventional choice we described above with $f=1/2$,  spins have a 'vision-cone' along $\bold{v}=(1,1)$: $\sigma=\pm1$ look at $\sigma\bold{v}$ with a higher interaction coupling.  
In  \cite{seara2023}, a constant uniform   vector $\bold{v}$ is introduced in the Ising couplings regardless of their state, globally driving the system. Our framework differs qualitatively as such bias is not global but dictated by the local spin state. 

For $f=1/4$, we made the choice of assigning to $\sigma=+1$ spins  a stronger coupling with the spin sitting at their top site than with the three remaining ones. Then, $\sigma=-1$ spins have a  stronger coupling with the other three remaining spins, the one at its bottom, right and left. The opposite applies for $f=3/4$. The last remaining possibility following this convention in the square lattice, is $f=0$ (or $f=1$). In this case a $\sigma=+1$ spin  interacts with a coupling $J$ with all its four neighbours, while a $\sigma=-1$ interacts with a coupling $J+\kappa$. This latter case is equivalent to adding an external field of strength $\kappa$. Lattices of higher connectivity offer more possible values of $f$. 

In the present work we only consider ferromagnetic couplings, meaning $\kappa>0$. We postpone the study of eventually competing antiferromagnetic bonds, $\kappa<0$, to a future investigation. 
Although here the model is defined in 2D, extensions to higher dimensions are straightforward. 

\subsection{Mean-field approach}\label{sec:mf-approach} 
We consider now a fully-connected network of $N$ spins for which the symmetric part of the coupling matrix is $J_{ij}^s = J/N$ for any $i\neq j$, with $J>0$. 
Given the definition of $\rho_{ij}$ in Eq. (\ref{eq:VC_FC}) for vision-cone like interactions, the evolution equation of the system's magnetization reads (see Appendix \ref{sec:mf-vc}),
\begin{equation}\label{eq:dyn_m_vc}
    \frac{d m}{d t} = -m + \tanh(Km) + \frac{1}{2}Kqm[1+m(1-2f)]\text{sech}^2(Km),
\end{equation}
where again $K = \beta J$ and $q = \kappa/J$ is the ratio between the non-reciprocal coupling and the reciprocal one.  $f$ represents the fraction of bonds for which the coupling of a spin $+1$ is larger, and comes from the specific choice of $\rho_{ij}$. 
In a fully connected network, the non-reciprocal coupling splits the system into two:  the fraction $f$ of spins in the system  to which $\sigma_i=+1$ interacts the most, and the remaining ones, a fraction $1-f$ of the bonds for which the coupling remains $J$. In this topology, the vision cone looses the geometric meaning it has in a finite dimensional lattice  shown in  Fig. \ref{fig:sketch}. 
 This interaction is naturally non$-$reciprocal: A spin $\sigma_i$ may have spin $\sigma_j$ in its "vision field", but $\sigma_i$ may not be in $\sigma_j$'s. 
Note how, if we set the non--reciprocal coupling to zero, $q = 0$, Eq. (\ref{eq:dyn_m_vc}) recovers the Ising model's evolution equation of $m$ in the mean--field picture, $\dot{m} = -m + \tanh Km$, which has a critical point at $K_c = 1$.

\begin{figure}[h!]
    \centering
    \includegraphics[scale = 0.65]{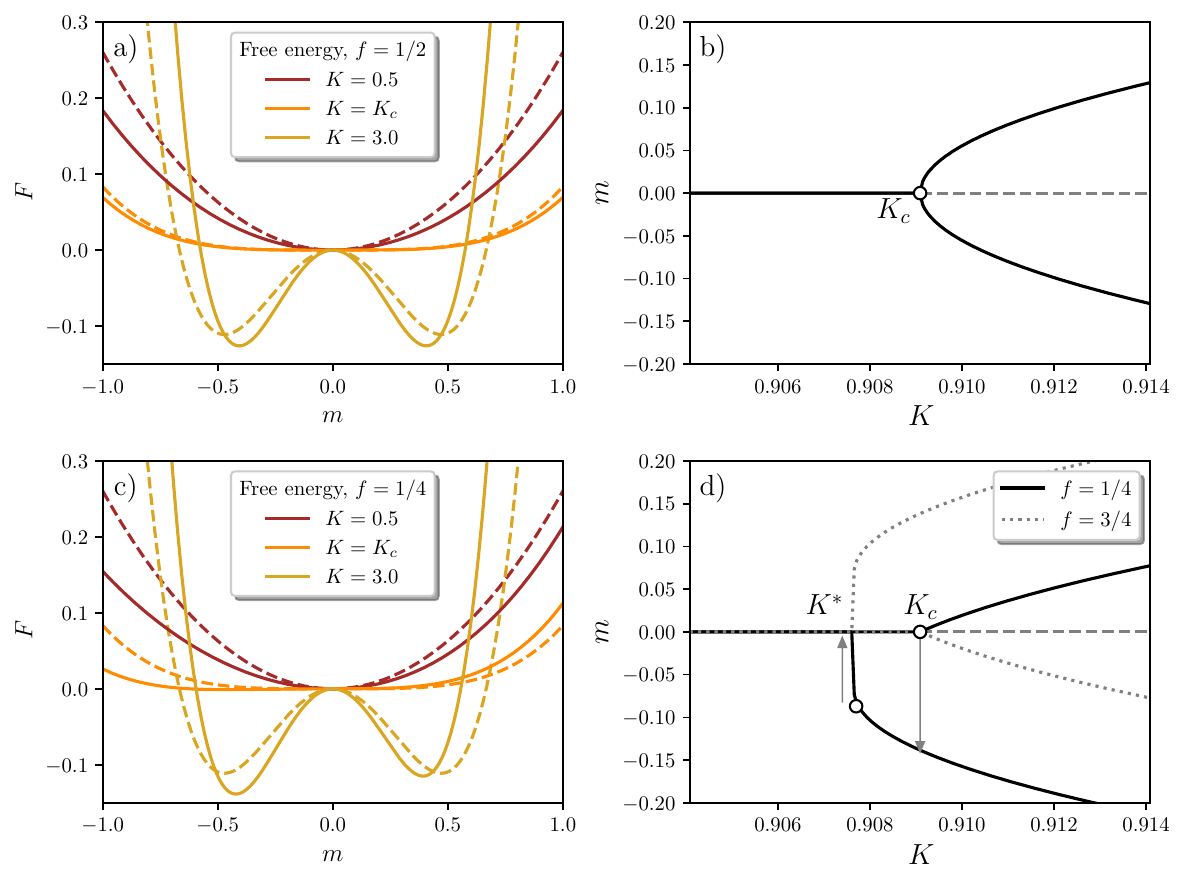}
    \caption{a) Mean-field free energy $F$ for a set of values of $K$ across the critical value $K_c$, at fixed $q = 0.5$ and $f = 1/2$. b) Stationary values of $m$ corresponding to the minima of $F$ shown in panel a). c), d) Same as a), b), now for $f=1/4$. 
    It is clear from the left panels how the free energy in the non-reciprocal models differs from the original mean-field Ising model (for which $K_c=1$), shown in dashed lines. }
    \label{fig:fef_and_lsc}
\end{figure}

In general, Eq. (\ref{eq:dyn_m_vc}) lacks   $m\rightarrow-m$ symmetry as long as $f \neq 1/2$. For $f = 1/2$, the steady magnetization has two symmetric indistinguishable branches. Instead, for $f \neq 1/2$, one gets two distinguishable branches, corresponding to the local and global minima of a mean-field free energy function, as discussed in more detail below. 
As shown in Eq. (\ref{eq:F_MF}), the time evolution of the order parameter Eq. (\ref{eq:dyn_m_vc}) can be rewritten as the relaxation of a function $F$
\begin{equation}
  \frac{d m}{d t} =- \frac{\partial F}{\partial m} 
\end{equation}
which,  after an expansion for  small $m$ reads,
\begin{align}\label{eq:fef_vc}
    F = &-\frac{1}{2}\left[K\left(1+\frac{q}{2}\right) -1\right]m^2 - \frac{1}{6}Kq(1-2f)m^3 +\frac{1}{12}K^3\left(1+\frac{3q}{2}\right)m^4 + \mathcal{O}(m^5).
\end{align}
This function  plays the  role of a Landau free energy in mean-field theory, and it is represented in Fig. (\ref{fig:fef_and_lsc}). 
For $f \neq 1/2$, the $m^3$ term in the Landau expansion, responsible for metastability, becomes non-zero, thus breaking the $m \rightarrow -m$ symmetry.
The symmetry is recovered when $f = 1/2$, as $F$ becomes then an even function of $m$. This is expected as, in this case, spins up or down interact with complementary fractions of the system. In this case, the impact of $q$ is to shift the critical point of the otherwise standard Ising model to a value $K_c=1/(1+q/2)$, that now depends on the non-reciprocal parameter $q$.   
Once $f\neq 1/2$, the system becomes biased, as a bigger (or smaller) fraction of spins is prioritised depending on the state of the spin considered.  For instance, $f=1/4$ gives a larger weight to the bonds connecting  a spin $\sigma_i=-1$ to its neighbours than the ones connecting a spin $+1$, resulting   in a deeper minimum of $F$ for negative values of $m$. This is reminiscent to the role played by an external magnetic field, which also triggers a discontinuous transition. However, an external field would show up in the free energy with a linear, temperature independent, term.  As illustrated in Fig. \ref{fig:fef_and_lsc} d), the $m^3$ term in the mean-field free energy generates a secondary minimum at $m>0$, i.e. a metastable state. 
%

%

The limit of stability, $K^{*}$, of the stable branch can be computed and gives,
\begin{equation}
    K^* = \frac{1}{2}\frac{1}{1+q/2}\left[1+ \sqrt{1 - \frac{3}{4}\frac{1+q/2}{1+3q/2}q^2(1-2f)^2}\right].
\end{equation}
Again, for $f=1/2$, we recover $K^*=K_c$, back to the symmetric situation. 
%
%
Note that  in the limit $q \rightarrow 0, K_c \rightarrow 1$, smoothly recovering the critical temperature of the mean--field Ising model.  
The reflection symmetry of the free energy functional reflects how the term $(1-2f)m^3$ makes the first and second order transitions exchange each other under the transformation $f \rightarrow 1-f$. Even though $K^*$ depends on the non--reciprocal parameter $q$, it is $f$ that controls the coexistence of a first and second order transition, and gives rise to a hysteresis cycle. For instance, in Fig. (\ref{fig:fef_and_lsc}, d), if the system is prepared at $K<K_c$ and the temperature is lowered, the state $m = 0$ will loose its stability exactly at $K = K_c$, and then spontaneously become negatively (positively) magnetized for $f= 1/4$ ($f = 3/4$). However, for $f = 1/4$ , if the system is prepared at a $K > K_c$ in the $m>0$ branch, and the temperature is increased, the  positively magnetized state will now be metastable until $K = K_c$. On the other hand, the $m<0$ branch is stable up to the limit of stablity $K=K^*$. Once $K<K^*$, the system will jump to a non--magnetized state $m = 0$. 

In the following sections we will see how the loss of the inversion symmetry plays a fundamental role also in finite dimensions $d$.

\subsection{Vision Cones with $\mathbb{Z}_2$ symmetry in $d = 2$.}\label{sec:finited_z2}
   
We now consider the model in a square lattice defined by Eq. (\ref{eq:rhoVC}) and focus first in the $\mathbb{Z}_2$ symmetric case ($f = 1/2$). 

One can  derive from Eq. (\ref{eq:finite_d_mf_evo}) a continuum description of the dynamics at coarse-grained scales. 
To do so, we set the symmetric term of the coupling matrix, $J_{ij}^s$, as the usual nearest-neighbour coupling of the Ising model in $d$ dimensions: $J_{ij}^s = J$ if $i$ and $j$ are nearest neighbours and $J_{ij}^s = 0$ otherwise. Then, using 
that $\delta_{(\sigma_i,\pm)} = (1\pm \sigma_i)/2$, and identifying  $(m_{i+\hat{e}_{\alpha}}-m_i)/a \rightarrow \partial_{\alpha} m_i$, the dynamic equation for the coarsegrained magnetisation field $m(\textbf{x},t)$ reads (see Appendix \ref{sec:app_e} for details),
\begin{align} \label{eq:continuum_vc_finite_d}
        \frac{\partial m(\textbf{x},t)}{\partial t} = &-(1-d\beta(2 J + \kappa))m(\textbf{x},t)- \frac{1}{3}(d\beta)^3(2J + \kappa)^3 m(\textbf{x},t)^3  \notag \\
        &+ a^2 \beta J \nabla^2 m(\textbf{x},t)+ \frac{1}{2}a \beta \kappa (m(\textbf{x},t)\textbf{v}\cdot \nabla)m(\textbf{x},t),
    \end{align}
where, again, $\textbf{v} = (1,1)^T$. 
 The uniform terms come from the Landau-like free energy, and  diffusion comes from the usual Ising (reciprocal)  coupling between first neighbours. The term $(m\textbf{v}\cdot \nabla) m$ is of an out-of-equilibrium nature. It has the form of a \textit{self}-advection of the field in the direction prescribed by our choice of non-reciprocal coupling, given by $\bold{v}$. 
 Such \textit{self-}advection term is commonly found in field--theories of active matter \cite{cates2019active}, such as the Toner--Tu theory of flocking \cite{tonertu1998}, or continuum descriptions of  XY spins with vision cone interactions  \cite{rouzaire2024non}. It is important to note that, while these latter theories involve vector fields, our theory Eq. (\ref{eq:continuum_vc_finite_d}) is scalar. Whereas the direction of \textit{self-}advection in continuum models like  Toner--Tu's is provided by the order parameter itself, 
  in the present model it is fixed and provided by the vision--cone chosen. 

The two first terms of the right hand side (RHS) of Eq. (\ref{eq:continuum_vc_finite_d}) imply that - in the mean field picture - the $f = 1/2$ admits the usual description with an effective coupling constant $\tilde{J} = J + \kappa /2$. This means that the critical inverse temperature $K_c$ separating the paramagnetic and the ferromagnetic phases will now depend on $\kappa$, as we  already saw in Sec. \ref{sec:mf-approach}.


To put the mean-field predictions into test and explore how non-reciprocal interactions affect the collective behaviour of the Ising model, we perform simulations in a $d = 2$ square lattice with periodic boundary conditions using Glauber dynamics. 
  
\begin{figure}[h!]
    \centering
    \includegraphics[width=1.0\textwidth]{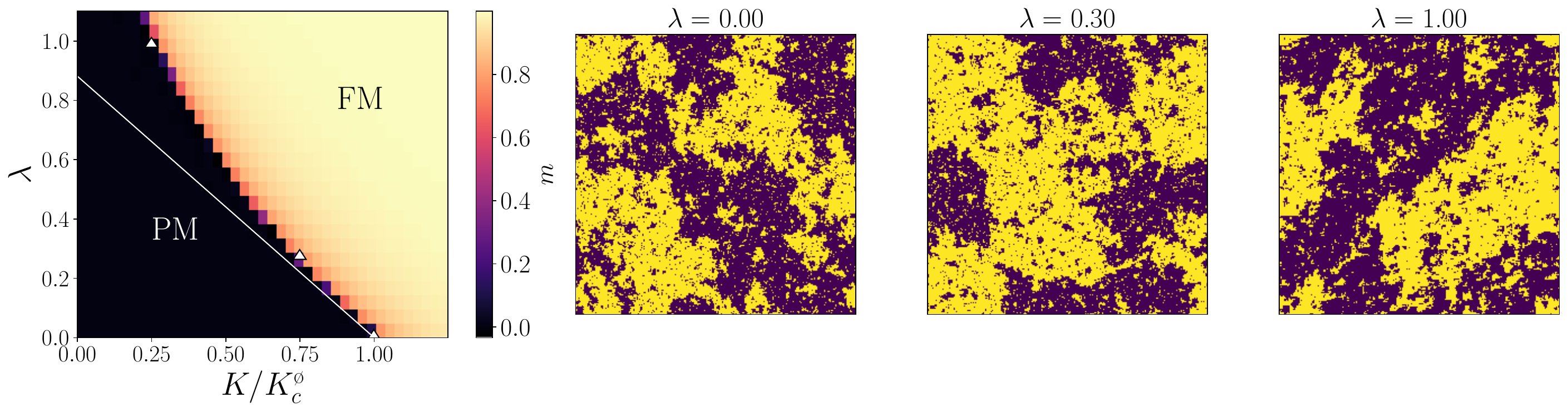}
    \caption{Phase diagram in the $K=\beta J, \lambda = \beta \kappa$ plane of the vision-cone model for $f = 1/2$. Simulations have been carried in a square lattice of size $L\times L = 100^2$, averaged over $ 100$ different realizations. The white line is the mean-field prediction of the correction of the critical line, $\lambda/2K_c^{\text{\o}}=1-K/K_c^{\text{\o}}$, where $K_c^{\text{\o}}$ is the {\O}nsager's exact critical temperature of the Ising model in a square lattice, $K_c^{\text{\o}} = \ln(1+\sqrt{2})/2$. 
    A color map is associated to the value of the magnetization measured in steady conditions, showing a  paramagnetic (PM) and ferromagnetic (FM) phase. On the right, snapshots of a $L\times L = 250^2$ lattice near the critical point for different values of $\lambda$, indicated by  triangles in the phase diagram. Spins $+1$ are colored in yellow, spins $-1$ in purple. }
    \label{fig:pd-vc-f-0.5}
\end{figure}

We show in Fig. (\ref{fig:pd-vc-f-0.5}) the map of the stationary global magnetization  obtained from simulations  at different values of $K = \beta J$ and $\lambda = \beta \kappa$ (here we use $\lambda$ instead of $q=\kappa/J$ as in Sec. \ref{sec:mf-approach}, where it was used to study  small deviations from the mean--field picture). In the phase diagram Fig. (\ref{fig:pd-vc-f-0.5}),  $K$ has been divided by $K_c^{\text{\o}} = \ln(1+\sqrt{2})/2$ the Ising's critical temperature in $d = 2$ such that when we set $\lambda = 0$ the transition occurs at $K/K_c^{\text{\o}} = 1$. The white line is the mean-field prediction of the critical point. It is given by $\tilde{K}_c = \beta \tilde{J} = 1$, where $\tilde{J} = J + \kappa/2$. At this level, non-reciprocity just renormalises the coupling strength, pushing the transition to higher temperatures. 
%
Such prediction of the critical temperature works fine for small $\lambda$, $\lambda \lesssim 0.3$. For $\lambda \gtrsim 0.3$, the line separating the paramagnetic (PM) and ferromagnetic (FM) phase starts deviating from the mean-field prediction. 
In the absence of the reciprocal coupling, $K = 0$, no FM phase can be observed.

\begin{figure}[h!]
    \centering
    \includegraphics[width=1\textwidth]{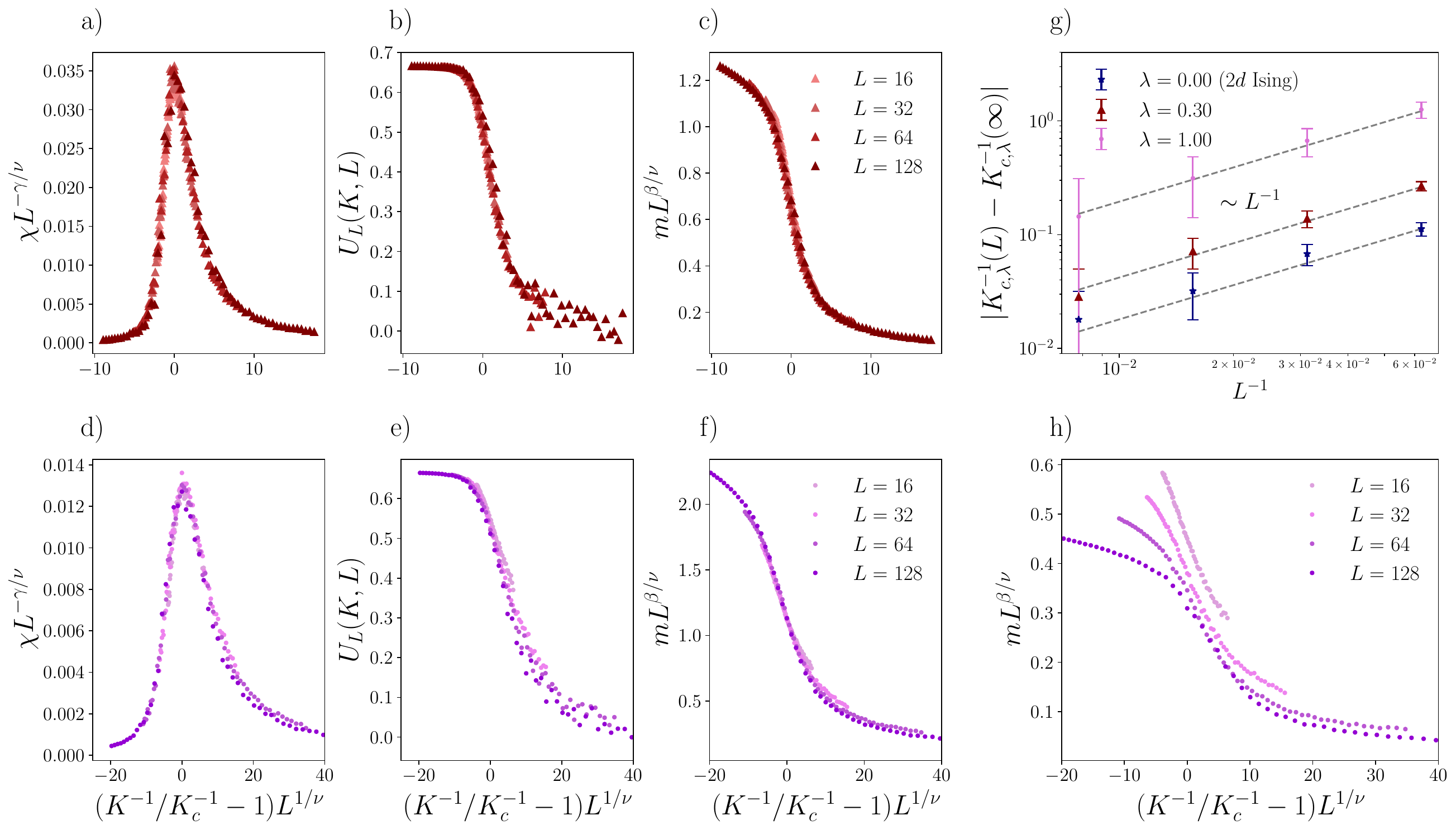}
    \caption{Finite size scaling of  the susceptibility,  the Binder parameter and the magnetization, for a range of sizes $L = 16,32,64,128$ 
    of the vision-cone model with $f = 1/2$ for $\lambda  = 0.30$ (a-c) and $\lambda = 1.0$ (d-f). The exponents $\nu$ and $\gamma$ have been set by hand to $\nu = 1$ and $\gamma = 7/4$ (the direct numerical estimations are indistinguishable from these values, see Table \ref{tab:table_exponents}); (g) dependence of the critical temperature $K_c^{-1}$ with the system size for $\lambda = 0.0$, $\lambda = 0.30$  and $\lambda = 1.00$. The gray dashed lines are $\sim L^{-1}$. Note how $K_{c,\lambda}^{-1}(\infty)$ has been first found by adjusting $K_{c,\lambda}^{-1}(L)$ as a function of $1/L$, taking then $K_{c,\lambda}^{-1}(\infty)$ as the value of the intersection with the axis $(1/L) = 0$; h) finite size scaling of the magnetization using  the Ising model's $d = 2$ critical exponents $\beta = 1/8$ and $\nu = 1$. The curves do not collapse on top of each other, suggesting that the model does not belong to the Ising universality class. }
    \label{fig:fss-vc-f-0.5}
\end{figure}

To get deeper insight into the nature of such phase transition in the presence of non-reciprocal couplings, we measure its associated critical exponents numerically. 
%
To do so, we perform a finite size scaling analysis in different regions of the phase diagram close and far away from the mean--field prediction of the PM-FM transition critical line, namely $\lambda = 0$ (Ising), $\lambda = 0.30$ and $\lambda = 1.0$ (triangles in Fig. (\ref{fig:pd-vc-f-0.5})), see Fig. (\ref{fig:fss-vc-f-0.5}).

The critical temperature $T_c$ in an infinitely large system  relates to the critical temperature measured in a finite system of size $L$, $T_c(L)$,  through the scaling $|T_c(L) - T_c| \sim L^{-1/\nu}$ \cite{fisher_barber_72, cardy2012finite}. This scaling relation helps us with two tasks: finding $T_c$ and $\nu$. Regardless of $\nu$, as long as its  positive, $T_c(L) \rightarrow T_c$ as $L\rightarrow \infty$. Since in our case the critical temperature eventually depends on $\lambda$, we can write $T_{c,\lambda}(L) \rightarrow T_{c,\lambda}$ as $L \rightarrow \infty$. In order to be consistent with the notation used so far, we will use $K = \beta J$ instead of $T$. 
To  obtain $K_{c,\lambda}^{-1}$, we locate the critical inverse temperature $K_{c,\lambda}^{-1}(L)$ as the point in which the susceptibility $\chi_{\lambda}(K,L)$ is maximum at fixed $L$, $\chi_{\lambda}^* = \operatorname{max}\{\chi_{\lambda}\}_{K} =  \chi_{\lambda}(K_{c,\lambda}, L)$. Using the collection of $K_{c,\lambda}^{-1}(L)$ for different sizes and fixed $\lambda$, a linear regression of $K_{c,\lambda}^{-1}$ with $L^{-1}$ can be used to provide an estimation of $K_{c,\lambda}^{-1}$. Then, one can estimate the exponent $\nu$ through the scaling $|K_{c,\lambda}^{-1}(L)-K_{c,\lambda}^{-1}| \sim L^{-1/\nu}$ by representing $|K_{c,\lambda}^{-1}(L) - K_{c,\lambda}^{-1}|$ versus the system size $L$ in a log--log scale. Using the peaks of the susceptibility for different sizes at fixed $\lambda$, $\gamma/\nu$ is estimated from $\chi_{\lambda}^*(L) \sim L^{\gamma/\nu}$. The exponent $\gamma$ is then extracted using the value of $\nu$ previously obtained. Once the critical inverse temperature and the exponent $\nu$ are obtained, the exponent $\beta$ 
 can also be computed via the scaling relation $m(K_{c,\lambda}(L)) \sim L^{-\beta/\nu}$. This is done by simply representing the value of the magnetization at the inverse critical temperature for a given size $m(K_{c,\lambda}(L))$ against the system size in a log$-$log scale.

The results obtained from performing such finite size scaling across the transition for $\lambda = 0.3$ and $\lambda = 1$ are shown in Fig. (\ref{fig:fss-vc-f-0.5}, a-c) and Fig. (\ref{fig:fss-vc-f-0.5}, d-f),  respectively. The scaling of the critical temperature with system size $L$ is consistent with an exponent $\nu = 1$,  as seen in Fig. (\ref{fig:fss-vc-f-0.5}, g) for both $\lambda = 0.3, 1.0$. In Fig. (\ref{fig:fss-vc-f-0.5}, a, b, d, e) the collapse of the susceptibility and the Binder cumulant across the critical temperature are shown using exponents $\nu = 1$ and $\gamma = 7/4$, in agreement with the $d = 2$ Ising model universality class. The finite-size scaling of the magnetization is shown in Fig. (\ref{fig:fss-vc-f-0.5}, c, f) where the exponents used are respectively $\beta \approx 0.15$ and $\beta \approx 0.2$ at fixed $\nu = 1$. These deviate from the $d = 2$ Ising model's exponent $\beta = 1/8$. To further back up this result, we show in Fig. (\ref{fig:fss-vc-f-0.5}, h) how, when using $\beta = 1/8$, the magnetization curves are far from collapsing into a single curve for $\lambda = 1.0$. The numerical values of the exponents obtained through finite size scaling are shown in Tab. (\ref{tab:table_exponents}). While the values of $\nu$ and $\gamma$ agree with the 2D Ising model ones, $\beta$ does not. As the measurement of the exponent $\beta$ from our finite-size analysis is the most delicate one, in the following we turn our attention into the dynamics of the model to provide further evidence of these estimations of the exponent. 

\begin{table}[h!]
    \centering
    \caption{Critical exponents obtained from simulations of the 2D Ising model with and without non-reciprocal couplings (for different values of $\lambda$).}
    \vspace{0.3cm}
    \begin{tabular}{cccc}
        \toprule
        \textbf{Exponent} & $\lambda = 0$ (Ising) & $\lambda = 0.3$ & $\lambda = 1.0$ \\
        \midrule
        $\nu$ & $1.04\pm 0.07$ & $0.94 \pm 0.07$ & $0.95\pm 0.06$ \\
        $\gamma$ & $1.74 \pm 0.04$ & $ 1.75 \pm 0.06$ & $1.73\pm 0.07$ \\
        $\beta$ & $0.13 \pm 0.04$& $0.15\pm 0.04$ & $ 0.21\pm 0.04$ \\
        $z$ & $1.97 \pm 0.04$& $1.97 \pm 0.08$ & $2.0 \pm 0.1$ \\
        \bottomrule
    \end{tabular}
    \label{tab:table_exponents}
\end{table}

\subsubsection{Short--time dynamics}

The relaxation of the magnetization after a quench from a fully ordered state, $m_0 = +1$, to a critical point, scales as \cite{janssen1989new, calabrese2006critical}
\begin{equation}\label{eq:short-time}
    m(t) \sim t^{-\beta/(\nu z)},
\end{equation}
%
where  $z$ is the dynamic critical exponent (see Sec. \ref{sec:4}). Thus, following the short-time relaxation of $m$ at the critical point, one can estimate $\beta/(\nu z)$ \cite{albano2011study}. The dynamic exponent $z$ is determined independently from the analysis of the coarsening dynamics in the following section.   
 We can, hence, compare the critical exponent $\beta$ obtained from the (static) finite-size scaling analysis for different values of $\lambda$ with the short-time dynamics results. 
We start from a fully ordered initial condition with all the spins pointing up, and let the system relax with a set of temperatures close to the critical temperature.

\begin{figure}[h!]
    \centering
    \includegraphics[scale = 0.35]{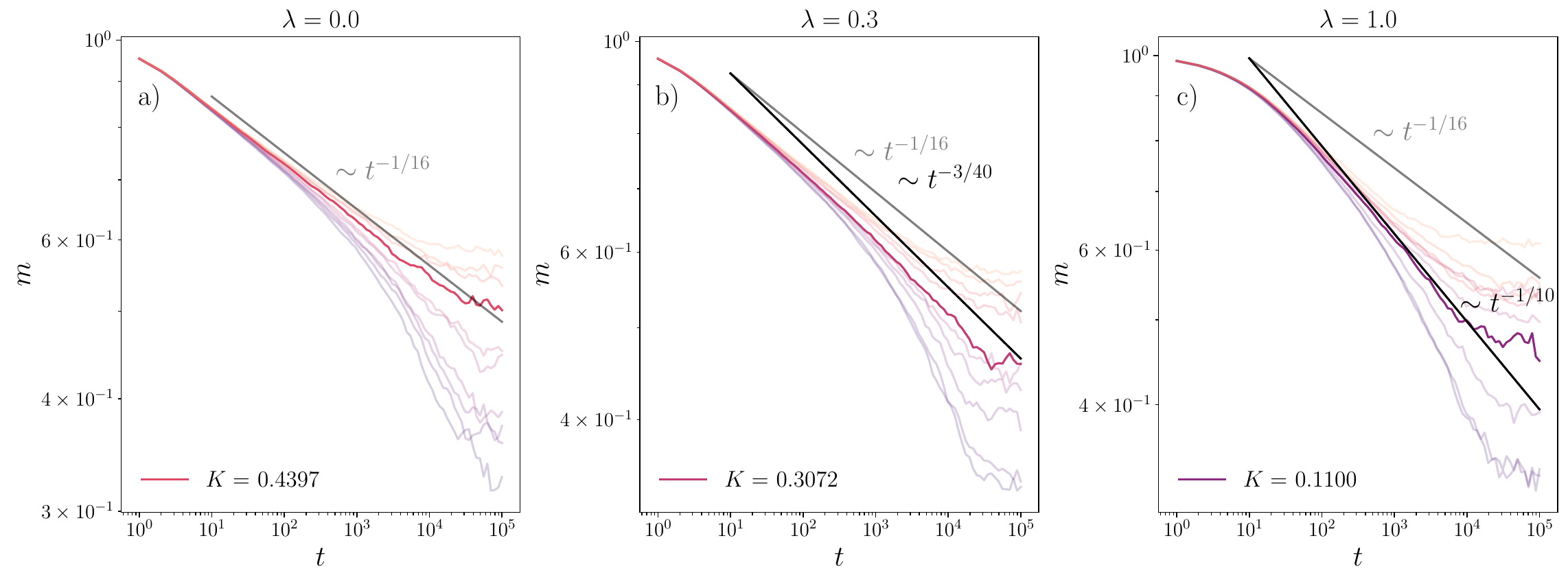}
    \caption{Relaxation of the magnetization of an initially ordered lattice with spins pointing up, $m_0=+1$ for 10 different temperatures around the critical point - for finite size $L$ - for a) $\lambda = 0$, the $d=2$ Ising model; b) $\lambda = 0.3$ and c) $\lambda = 1$ for a system of size $L\times L = 256^2$. 
    }
    \label{fig:std_comparison}
\end{figure}

We show in Fig. (\ref{fig:std_comparison}) the time evolution of the magnetization following a quench from a $m_0=+1$ state to different values of $K$. The decay of the magnetization follows  a power law at criticality, at the values of $K$ for which the susceptibility exhibits a maximum for this system size. 
The dark lines correspond to the decay $m\sim t^{-\beta/(\nu z)}$ using the static exponents $\beta$ and $\nu$ obtained from the finite size scaling, while $z=2$ is the value extracted from the analysis of the dynamics in Sec. (\ref{sec:4}), that does not change with $\lambda$. The value of the exponents is given in  Tab. (\ref{tab:table_exponents}). As one can appreciate in Fig. (\ref{fig:std_comparison}, a-c), the larger  $\lambda$ is, the faster the relaxation is, in agreement with our previous finding. 
Since the exponent of the decay  $m\sim t^{-\beta/(\nu z)}$ does change with $\lambda$, as the data Fig. (\ref{fig:std_comparison}) shows, the critical exponent $\beta$  also does change, going beyond the 2D Ising model universality class.

 \subsection{Vision Cones lacking $\mathbb{Z}_2$ symmetry in $d = 2$.} \label{sec:finited_no_z2}
 
 Let us consider now a case for which the mean-field free energy functional in Eq. (\ref{eq:fef_vc}) is not symmetric, that is, $f \neq 1/2$. For $f = 3/4$, the non-reciprocal, vision-cone  interactions, are defined by 
 \begin{equation}\label{eq:rhoVC_f_0.75}
         \rho_{ij} = \delta_{\sigma_i,+}(\delta_{i+\hat{e}_x,j} + \delta_{i+\hat{e}_y,j}+\delta_{i-\hat{e}_x,j}) +  \delta_{\sigma_i,-}\delta_{i-\hat{e}_y,j}
 \end{equation}
 In this case spins 'up' ($+1$) add extra coupling $\kappa$ to its right, left and top bonds to the overall, isotropic and reciprocal coupling $J$, while spins 'down' ($-1$) only add extra coupling to the bottom bond. Note that such non-reciprocal coupling cannot be generalized to an interaction along a single vector $\textbf{v} =  \lambda_x \hat{e}_x + \lambda_y \hat{e}_y$, as it was possible in the previous definition.

 Following now Eq. (\ref{eq:finite_d_mf_evo}), the continuum description at coarse-grained scales of the magnetization field $m(\textbf{x},t)$ becomes
\begin{align}\label{eq:continuum_vc_finite_d_f_3/2}
     \frac{\partial m(\textbf{x},t)}{\partial t} = &-(1-2\beta(2 J + \kappa)) m(\textbf{x},t) + \beta \kappa m^2(\textbf{x},t) -\frac{1}{3} (2\beta)^3(2J+\kappa)^3m^3(\textbf{x},t) \notag \\
     & + a^2\beta J \nabla^2 m (\textbf{x},t) + \frac{1}{2}a^2\beta\kappa(1+m)\partial_x^2 m(\textbf{x},t) + \frac{1}{2} a \beta \kappa m \partial_y m (\textbf{x},t).
 \end{align}
Eq. (\ref{eq:continuum_vc_finite_d_f_3/2}) differs from Eq. (\ref{eq:continuum_vc_finite_d}) in various ways: 
a new term $m^2(\textbf{x},t)$ enters the dynamical relaxation of $m(\textbf{x},t)$ as a consequence of the breakdown of the inversion symmetry. 
Similarly to the $f=1/2$ case, non-reciprocity gives rise to a non-equilibrium self-advection term, now along the direction $\hat{e}_y = (0,1)^T$. However, in this case, another unusual term appears: a diffusive term, with state dependent diffusion coefficient along the direction $\hat{e}_x = (1,0)^T$, $D_{x}(m) \sim (1+m(\textbf{x},t))/2$. 

\begin{center}
    \begin{figure}[t]
        \includegraphics[scale=0.45]{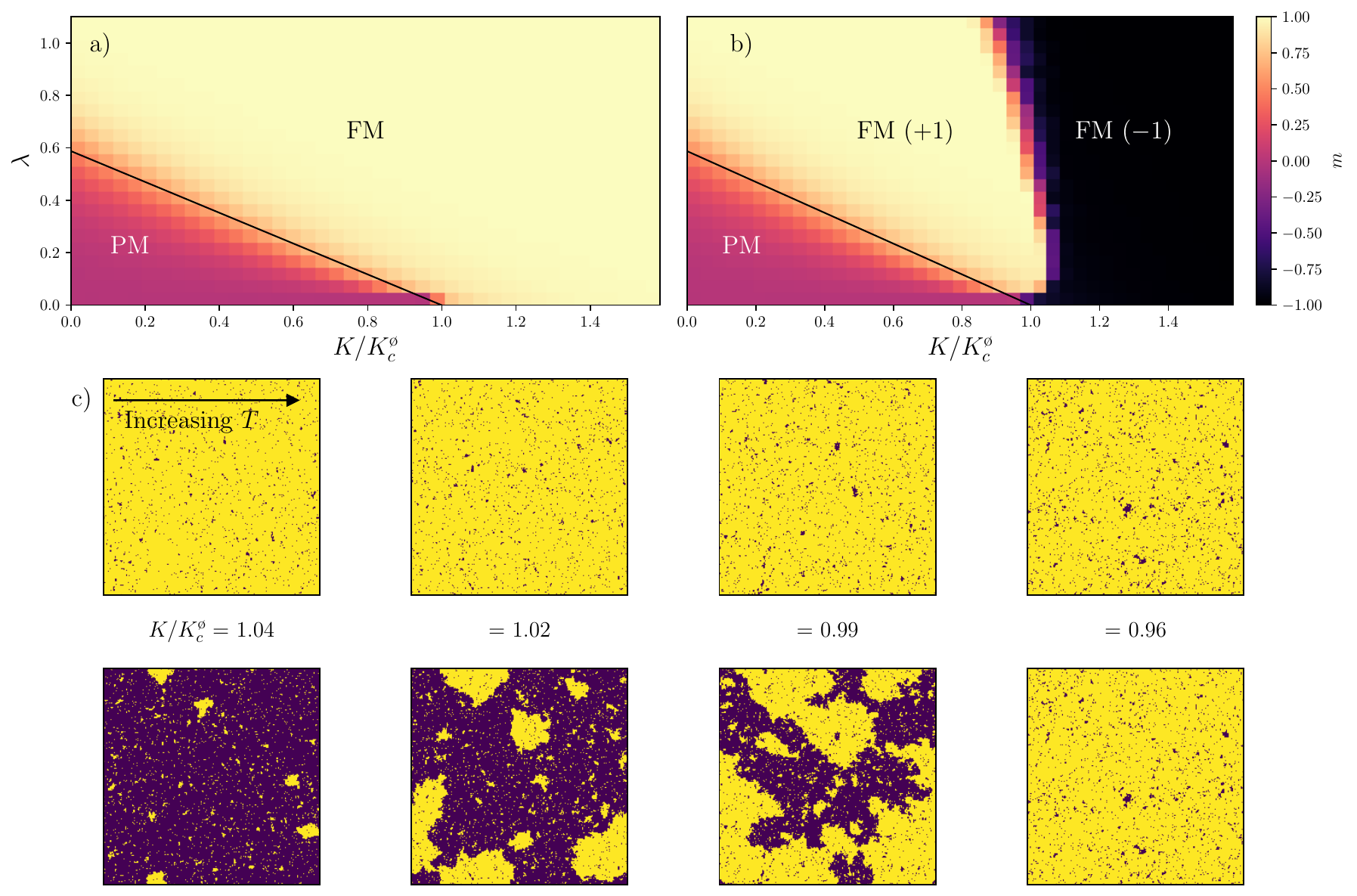}
         \caption{Discontinuous and continuous transition in the $f=3/4$ vision cone. (a) Color map of the stationary global magnetization obtained following a quench from an initially ordered, $m_0 = +1$, lattice of size $L\times L = 100^2$ spins 
         for the $f=3/4$ vision--cone model, averaged over 100 different realizations 
         The black line represents the mean--field prediction of the critical line, $3\lambda/4K_{c}^{\o} = 1 - (K/K_{c}^{\o})$, where again $K_c^{\o}$ is the $d = 2$ critical inverse temperature, separating a PM and a FM phase; (b) \textit{idem} now for an initially ordered state $m_0 = - 1$; (c) snapshots of the system evolved through subsequent temperature increases from an  initially ordered lattice, $m_0 = +1$ top row and $m_0 = -1$ bottom row, of size $L = 250$ spins for  $\lambda = 0.2$. Each frame represents a steady-state configuration of the lattice. The temperature is increased slowly, letting the lattice relax before increasing the temperature  again.}
        \label{fig:vc-f-0.75-pd}
    \end{figure}
\end{center}

The phase diagram of the model for $f=3/4$ is shown in Fig. (\ref{fig:vc-f-0.75-pd}), now differentiating the $m>0$ and $m<0$ branches. Indeed, the system exhibits hysteresis, a signature of a discontinuous phase transition, as expected from the mean-field picture.  Fig. (\ref{fig:vc-f-0.75-pd}.a) shows the steady magnetisation obtained from the relaxation of a $m_0=+1$ configuration in  simulations, while  Fig. (\ref{fig:vc-f-0.75-pd}.b) shows the counterpart, obtained from a $m_0=-1$ initial configuration. 
The mean-field prediction of the critical line $K_c(\lambda)$ is in reasonable agreement with the numerics. Within the FM phase, the breakdown of spin reversal symmetry translates into metastability and the emergence of a discontinuous phase transition. Starting from a $m<0$ initial state, the system relaxes to a FM state with $m<0$ up to $K\approx 1$ beyond which it 'jumps' to the stable $m>0$ state (in yellow). In mean-field terms,  $m<0$ states correspond to a local minimium of the free energy, while $m>0$ states correspond  global minima. 
Indeed, for $f=3/4$, there is a preference for positively magnetized states in the FM phase. A spin pointing up interacts with an effective coupling constant $\tilde{J} = J + 3\kappa/4$, while for spins pointing down  $\tilde{J} = J + \kappa/4$.  At low enough temperatures, the system can get trapped in the metastable branch. This can be again understood from the mean-field picture:  at low temperature, the energy barrier separating the two minima, the stable from the metastable branch, is big enough to prevent thermal fluctuations to overcome it in the simulation time scales. However, as temperature increases, the height becomes smaller and it is more likely for the system to jump from the metastable state into the stable one.  

A region of the phase diagram close to the limit of stability of the $m<0$ state (i.e. close to $K=1$) is portrayed in Fig. (\ref{fig:vc-f-0.75-pd}. c) for $\lambda$. 
A large enough domain of $+1$ spins in a sea of $-1$ spins can eventually nucleate and lead to $m>0$.  
Then, as the temperature increases, fluctuations are enhanced and it becomes easier to cross the energy barrier to nucleate a $+1$ domain. On the contrary, $m>0$ states, remain stable allover the FM phase.

As opposed to the $f = 1/2$ case, for $f = 3/4$ the system exhibits both a discontinuous and continuous phase transition, as predicted by the mean-field model. 
The 
 effective coupling constant of spins +1, corresponding to the stable branch, is
$\tilde{K} = \beta \tilde{J} = K + 3\lambda/4$. The critical mean-field line is thus given by  $3\lambda /4K_{c}^{\o} = 1 - (K/K_{c}^{\o})$, which seems to nicely agree with the emergence of  magnetized states as reported in Fig. (\ref{fig:vc-f-0.75-pd}). 
The  line separating  PM and FM states  does not grow asymptotically when $K \rightarrow 0$, but intersects the vertical axis ($K = 0$) at finite $\lambda$,  ($\lambda = 4K_{c}^{\o}/3\approx 0.59$ in mean-field), see Fig. (\ref{fig:vc-f-0.75-pd}). This means that strong enough non-reciprocal coupling $\lambda$ can induce FM order by itself, with no need of an underlying FM exchange coupling $J$. Beyond $\lambda\approx 0.6$ the system acquires a global magnetization although $K=0$. 
 
\begin{figure}[h!]
    \centering
    \includegraphics[scale=0.55]{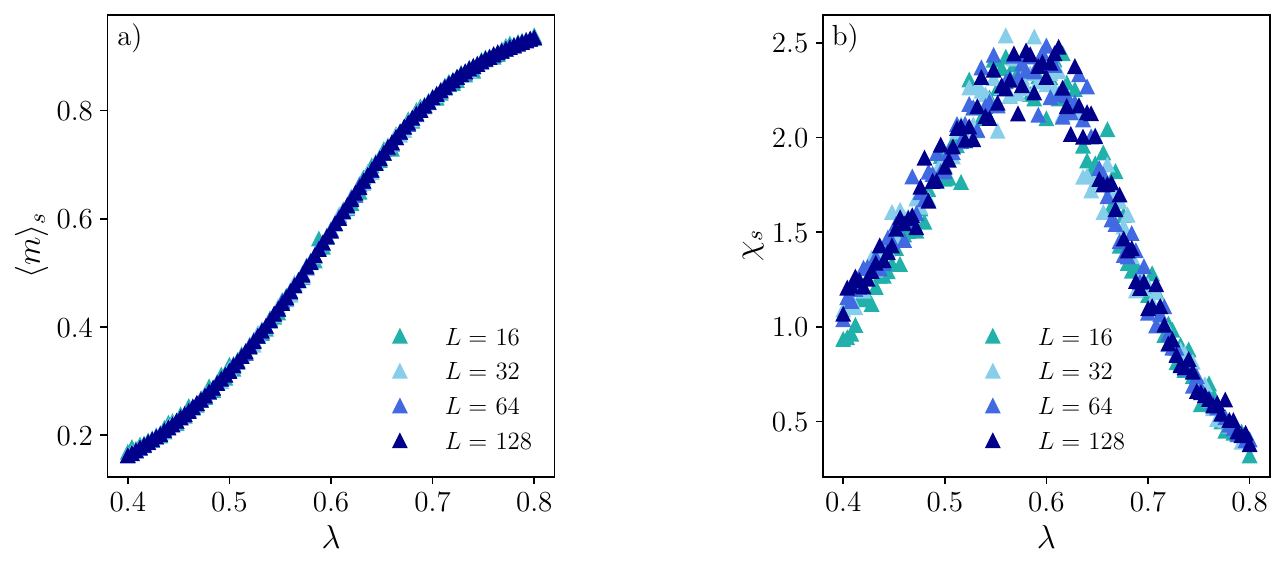}
    \caption{(a) Steady magnetization, $\langle m \rangle_s$ as a function of $\lambda$ 
    for $K=0$ in the $f=3/4$ vision--cone model. The system sizes used are indicated in the legend. 
    (b) Corresponding susceptibility, exhibiting a maximum close to $\lambda\approx 0.6$. 
    Neither the magnetization nor the susceptibility scale with system size $L$ in the absence of reciprocal coupling.}
    \label{fig:fss_f_0.75_K_0}
\end{figure}

In order to better grasp the behaviour of the system at $K=0$, we run simulations on square lattices of different  sizes, varying $\lambda$ across the mean-field critical line. The results are shown in  Fig. (\ref{fig:fss_f_0.75_K_0}). Systems of different size $L$ show similar magnetization curves. The associated susceptibility  exhibits a peak at $\lambda\approx 0.6$ for any system size. Such peak does not become more and more pronounced upon increasing $L$, as occurs close to a critical point. We thus conclude that it corresponds a  crossover between the PM-FM states indicated in the phase diagram Fig. (\ref{fig:vc-f-0.75-pd}) at $K=0$. Although mean-field theory predicts a critical line between such states,  we did not find any evidence of such behaviour in the square lattice simulations. 

\section{Coarsening dynamics}\label{sec:4}

We turn now into the analysis of the coarsening dynamics of the $f=1/2$ model following a  quench from an initially random configuration to an ordered state set by $K>K_c$. 
The aim of the analysis presented in this section is two fold. From one hand, understanding whether the presence of self-advection at the continuum level might modify the dynamic scaling properties of model A (or time dependent Ginzburg-Landau model). From another hand, estimating the dynamic exponent $z$, needed to confirm the value of the  exponent $\beta$ (measured from the finite-size scaling analysis)  using the short-time relaxation method discussed earlier. 


\subsection{Coarsening of Vision--Cones with $\mathbb{Z}_2$ symmetry}

The $f=1/2$ vision-cone interaction breaks isotropy, introducing a preferential direction $\textbf{v}= (1,1)^T$ along which FM interactions are stronger.  
This suggest to analyze the growth of spin-spin correlations (or domains) along the direction determined by $\textbf{v}$ and the direction perpendicular to it, provided by the following correlation functions in space and time
\begin{align}
        C_{\parallel} (n,t) &= \Big \langle \frac{1}{N}\sum_{i,j}\sigma_{i,j}(t)\sigma_{i+n, j+n}(t) \Big \rangle \label{eq:corr-vc-f-0.5-para} \\
        C_{\perp} (n,t) &= \Big \langle \frac{1}{N}\sum_{i,j}\sigma_{i,j}(t)\sigma_{i-n, j+n}(t) \Big \rangle \label{eq:corr-vc-f-0.5-perp},
\end{align}
where $\sigma_{i,j}$ denotes the Ising spin sitting on the node of coordinates $(i,j)$, $n$ being an integer number.


\begin{figure}[h!]
    \centering
    \includegraphics[scale=0.5]{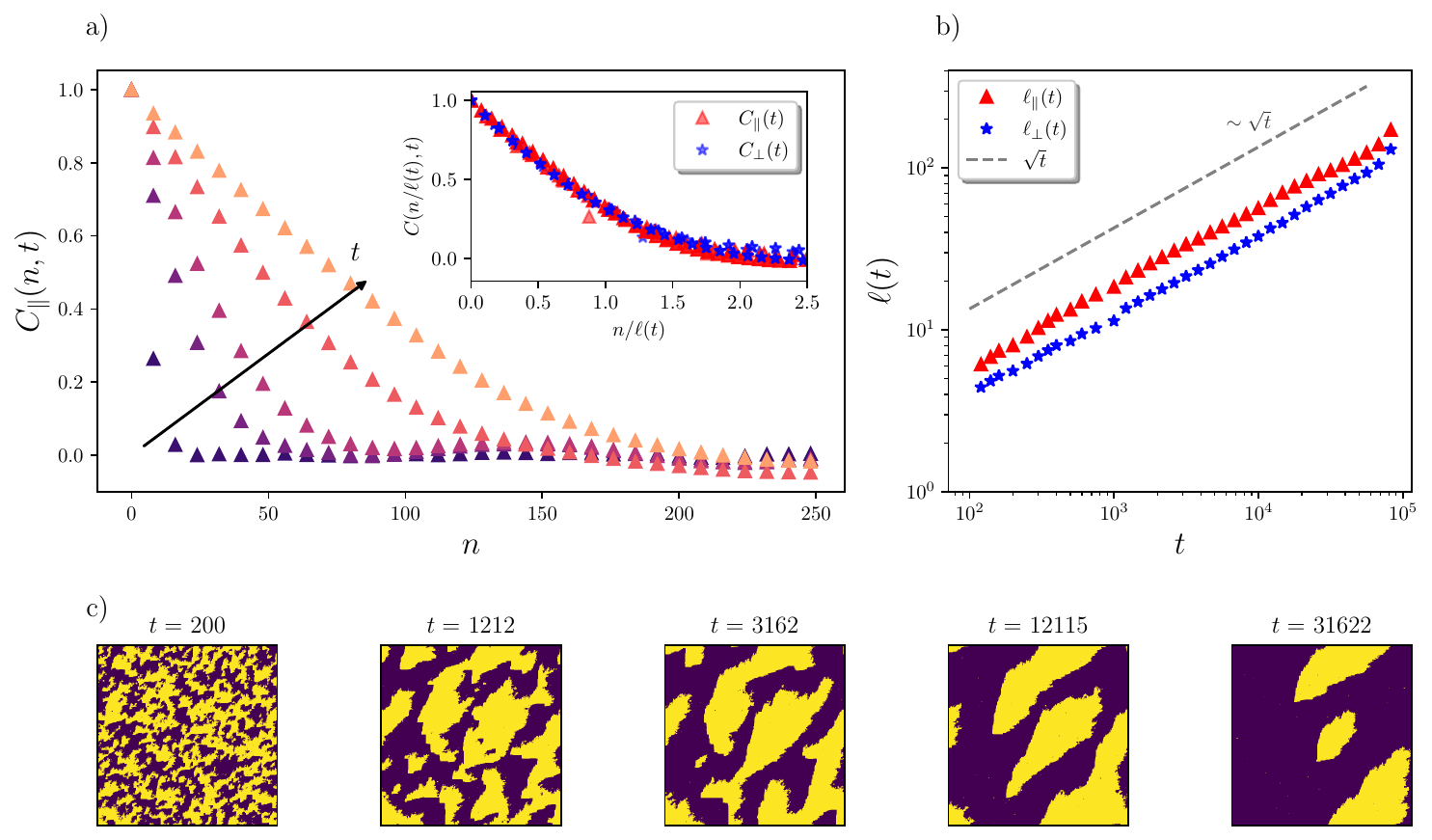}
    \caption{Coarsening dynamics of the vision--cone model for $f = 1/2$.  (a) Spatial decay of the correlation function along the direction of the vision cone, for different times after a quench  $t = 200, 1212, 3162, 12115, 31622$ (from blue to yellow) versus the distance $n$ for systems of size $L=500$. 
    The inset shows the collapse of the correlation curves along the direction of the vision cone and perpendicular to it at different times, once the lattice distance $n$ has been rescaled by the typical growing lengths $\ell_{\parallel,\, \perp}(t)$. 
 (b) Growth of the two characteristic lengths after a quench from an initially disordered state at $(K,\lambda) = (0.2,2.0)$. Such lengths have been obtained from $C_{\parallel,\perp}( \ell_{\parallel,\, \perp} ,t)\approx e^{-1}$ at a given time $t$. (c) Snapshots of the evolution of the system for different times following the quench. Although the domain growth is anisotropic, the dynamic scaling exponent in both directions remains $z\approx 2$. }
    \label{fig:coarsening_f_0.5}
\end{figure}

The spatial decay of $C_{\parallel}$ is shown in Fig. (\ref{fig:coarsening_f_0.5}) at different times following the quench. As time goes on, spatial correlations build up, and the associated typical length scales, or correlation lengths, increase. This translates the fact that ferromagnetic domains grow in time, as illustrated by the snapshots Fig. (\ref{fig:coarsening_f_0.5}, c). We further quantify such domain growth by extracting  two characteristic length scales  in the direction parallel, $\ell_{\parallel}(t)$, and perpendicular, $\ell_{\perp}(t)$,  from the correlations by $C_{\parallel,\perp}( \ell_{\parallel,\, \perp} ,t)\approx e^{-1}$. As shown in  Fig. (\ref{fig:coarsening_f_0.5}), domain growth is biased. Domains are elongated, growing more along the direction set by the vision--cone, see Fig. (\ref{fig:coarsening_f_0.5}, c). 
However, the scaling of both $\ell_{\parallel}(t)$ and $\ell_{\perp}(t)$ seem to agree with  the Ising model scaling in $d = 2$:
\begin{equation}
\ell_{\parallel,\, \perp}(t)\sim t^{1/z},\ z=2\,
\end{equation}  
also in the presence of non-reciprocal coupling. Thus, going back to the analysis of the short-time dynamics, the domain growth is consistent with an exponent $\beta$ increasing with $\lambda$. 



\subsection{Coarsening using a drop}

Following a quench from a disordered  to an ordered state, the phase  ordering kinetics  proceeds via the growth of magnetic domains. The asymptotic growth law governing the dynamics at long times (yet smaller that the equilibration time in a finite-size system) can alternatively be studied by tracing the evolution of a spherical magnetic domain, or {drop},  with a given magnetization  in a sea with the opposite orientation \cite{bray2002theory, di2024off}. In the following, we analyse the dynamics of a spherical drop made of spins $+1$, surrounded by spins $-1$, to provide yet another independent estimation of the value of  the growth exponent $z$. 

In the  $T \rightarrow 0$ limit, the shrinking dynamics of a spherical domain immersed in a sea with opposite magnetization is analytically treatable and provides substantial insights into the phase ordering kinetics. The fate of a finite drop in an environment with opposite magnetisation (the majority phase)  is  to shrink isotropically until it vanishes, reducing its surface  (the radius), that carries an excess interface energy. The time evolution of the radius of such drop over time, $R(t)$, can be explicitly computed for model A \cite{bray2002theory} (see Appendix \ref{sec:app_f}). It can be shown that the time it takes for the drop to shrink until it disappears - the shrinking time, $\tau_s$- scales with the initial radius of the drop, $R_0$, as 
\begin{equation}
\tau_s \sim R_0^z
\end{equation} 
where $z = 2$. The dynamical evolution of the radius of the shrinking drop $R(t)$ can be directly related to the global magnetization of the lattice. For a bubble of spins pointing up in a sea of spins pointing down, the magnetization evolves first linearly, $m(t) = m_0 -At$, before saturating to the steady state $m = -1$ as $t$ increases. The initial linear decay of $m$ can be used to extract the shrinking time $\tau_s$ as the time for which $m(\tau_s) = -1$, giving $\tau_s = (1+m_0)/A$. The dependence of the the shrinking time on the drop's initial radius $R_0$ is thus encoded in $m_0$. 

\begin{figure}[h]
    \centering
    \includegraphics[width=1.0\linewidth]{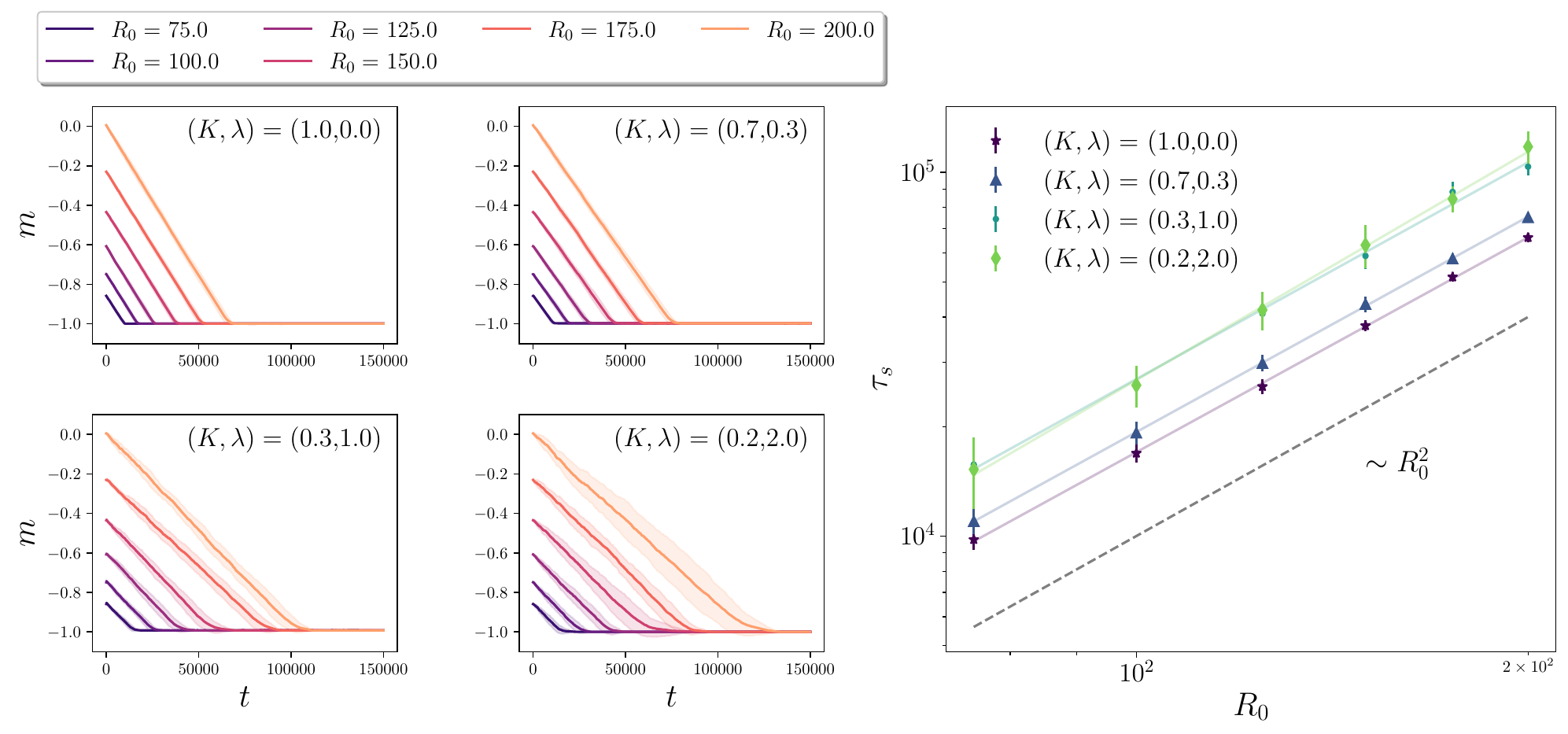}
    \caption{Left: Time evolution of the magnetization in systems of size  $L = 500 $ initially prepared with a spherical domain of spins $\sigma = + 1$ of radius $R_0$ (given in the legend) surrounded by spins $\sigma = -1$ for the different values of $(K,\lambda)$ indicated in the legend. Right: scaling of the shrinking time $\tau_s$ with the initial size of the drop $R_0$ for the set of values $(K,\lambda)$ in the left panels.  }
    \label{fig:drop_coarsening}
\end{figure}

We perform such an analysis for our model with $f=1/2$. We  prepare an initial configuration with a FM drop of $+1$ spins in a sea of $-1$ spins and let it relax towards its steady state (with negative magnetisation).  In our simulations at finite temperature, thermal fluctuations prevent the lattice from reaching a perfectly ordered steady state $m=-1$. However, for $T<T_c$, the initially prepared drop with $m=1$  indeed shrinks and the  global magnetization decays in time. 
From the initial linear decay of $m$ shown in Fig. (\ref{fig:drop_coarsening}) for different initial drop sizes, we extract the values of $\tau_s$ reported. 
The data is, again, consistent with the dynamic universality class of the Ising model with Glauber dynamics (non-conserved order parameter dynamics), namely,
\begin{equation}
\tau_s \sim R_0^z,\ z=2 \,.
\end{equation} 
The non-reciprocal parameter $\lambda$ appears thus in the prefactor of the scaling law, just as $K$ does in the reciprocal situation. Both the analysis of the coarsening dynamics and the shrinking of an unstable drop give compatible results, providing numerical evidence to the values of critical exponents reported in Table  (\ref{tab:table_exponents}).

\section{Discussion and conclusions}\label{sec:5}

We have presented a general framework to investigate spin models with non-reciprocal couplings, and shown how, formally, it affects the continuum description \`a la Ginzburg-Landau of the Ising model. We then focus our analysis on a specific non-reciprocal Ising model, mimicking a vision-cone. By defining the model on a fully connected network, we find that the phase diagram of the model at the mean-field level can be understood by means of an effective Landau-type free energy function. 
We then focus our attention on two type of vision cones, breaking or not the spin inversion symmetry. 
In the former case,  the model exhibits a discontinuous and a continuous phase transition, both in the square lattice and at the mean-field level. 
A thorough analysis of the second case that does not violate the Ising $\mathbb{Z}_2$ symmetry, reveals that the continuous PM-FM transition is not characterised by the same critical exponents of the 2D Ising model universality class: the exponent $\beta$ depends on non-reciprocity. 

The continuum description of both variants of the vision-cone model, feature non-equilibrium terms, in the form of self-advection and anisotropic 'self-enhanced' diffusion. Exploring the role of these terms in the large-scale physics of scalar field theories is surely of interest, as well as its connection with similar vectorial theories of flocking. As the framework presented here could be adapted to include mixtures (non-reciprocal interactions between different species), it would surely be of interest to investigate non-reciprocal Ising models with conserved order parameter dynamics, and the corresponding conserved dynamics at the field theory level. Overall the present work opens several new lines of research that we believe should shed light on the impact of non-reciprocal interactions on large-scale phenomena, and in particular on phase transitions and critical phenomena. 

\backmatter

\bmhead{Acknowledgements}
A. G. acknowledges AGAUR and Generalitat de Catalunya for financial support under the call FI SDUR 2023.
D. L. acknowledges MCIU/AEI and DURSI for financial support under Projects No. PID2022-140407NB-C22 and 2021SGR-673, respectively. 

\newpage

\begin{appendices}

\section{Extension to Multiple Species}\label{sec:app_a}
Many recent studies have considered non--reciprocal interactions between degrees of freedom of different species in multi--component systems
 \cite{agudo-canalejo,saha2020scalar, dinelli2023nonreciprocal, theveneau2013chase, suchanek2023irreversible, suchanek_23, tucci2024nonreciprocal, osat2023non, traveling_Marchetti, saha2022effervescent, guislain_mean_field, Guislain_2024_spin, avni2023non, Fruchart2021, lorenzana2024non, guislain2024far}. In such systems, spins $\sigma_{i}^{\mu}$ - not particularly Ising type -  of a given species $\mu = A, B, \dots$ , where $i = 1,\dots, N_{\mu}$ ($N_{\mu}$ is the number of spins of species $\mu$) interact with spins of species $\nu = A, B, \dots$, $\sigma_{j}^{\nu}$ (where again $j = 1, \dots, N_{\nu}$ and $N_{\nu}$ is the number of spins of spins of species $\nu$) with a coupling constant that only depends on the species involved, $J_{\mu\nu}$. One usually considers that spins of the same species, say $A$, interact reciprocally with coupling constant $J_{A}$, whereas spins of different species, say $A$ and $B$, interact in a non--reciprocal way, with couplings $J_{AB}\neq J_{BA}$. Frustrated interactions, e.g $J_{AB}=-J_{BA}$ generates cicles or run--away dynamics, as observed in a  variety of systems  such as the ones in refs. \cite{Fruchart2021, avni2023non, traveling_Marchetti, saha2020scalar, guislain_mean_field, Guislain_2024_spin, Guislain_disordered}.


\begin{figure}[h!]
    \centering
    \includegraphics[width=0.9\linewidth]{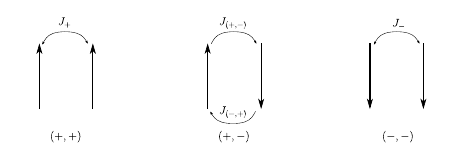}
    \caption{Sketch showing how the the state of spins can be interpreted as the species they belong to and how this can be used to introduce non--reciprocal interactions that mimic a mixture of two species of spins interacting.}
    \label{fig:enter-label}
\end{figure}

Our model, thus, can be used to mimic interactions between two species of spins, if the state of the spin is considered to be the species it belongs to. Following such identification, the interactions should depend on the state (or species) of both spins in a bond. 
This can be done by introducing a dependence on the state of both spins in a bond in the coupling matrix.
For instance, a very simply example would be setting the symmetric part of $J_{ij}$ to zero, $J_{ij}^{s} = 0$, and define 
\begin{equation}
\rho_{ij} = \delta_{(\sigma_i, \sigma_j)} + \delta_{(\sigma_i,+)}\delta_{(\sigma_j,-)} - \delta_{(\sigma_i,-)}\delta_{(\sigma_j, +)}
\end{equation} 
meaning,
\begin{align}
    J_{ij} = \kappa \left[\left(\frac{1+\sigma_i \sigma_j}{2}\right) + \left(\frac{\sigma_i - \sigma_j}{2}\right) \right],
\end{align}
so that spins of the species interact with coupling $\kappa$, $+1$ spins interact with $-1$ spins with coupling $J_{+-}=\kappa$ while $-1$ spins interact with $+1$ spins with coupling $J_{-+}=-\kappa$. 
For binary spin systems, the number of $+1$ and $-1$ spins, if interpreted as particles of two different species, a conserved dynamics of the Kawasaki type seems more adequate. 


\section{Master Equation and Spin Flip Dynamics}\label{sec:me-sfd}
Consider an observable $\mathcal{O}(\boldsymbol{\sigma})$, with no implicit time dependence (it is a measure of the configuration of the spins). Then, we know that, by definition
    \begin{equation}
        \langle \mathcal{O}\rangle_{t} = \sum_{\{\boldsymbol{\sigma}\}} \mathcal{O}(\boldsymbol{\sigma}) p(\boldsymbol{\sigma};t).
    \end{equation}
where   $p(\boldsymbol{\sigma};t)$ is the probability of findinf the system in a given microstate at time $t$.   If the rate of flipping spin $\sigma_i$ is $\omega_{i}(\sigma_i)$, the master equation for spin flips reads,
    \begin{equation}
        \frac{d}{d t} p(\boldsymbol{\sigma};t) = \sum_{i} \left[\omega(\sigma^i) p(\boldsymbol{\sigma}^i;t) - \omega(\sigma_i)p(\boldsymbol{\sigma};t)\right],
    \end{equation}
    where if $\boldsymbol{\sigma} = (\sigma_1,\dots, \sigma_i, \dots,\sigma_N)^T$, then $\boldsymbol{\sigma}^i = (\sigma_1,\dots,-\sigma_i, \dots,\sigma_N)^T$ after flipping spin $\sigma_i$, and, samewise, $\sigma^i = - \sigma_i$. Now, we can change $p(\boldsymbol{\sigma};t)$ by $\mathcal{O}(\boldsymbol{\sigma}) p(\boldsymbol{\sigma};t)$, so that, summing over all the configurations $\{\boldsymbol{\sigma}\}$,
    \begin{align}
        \sum_{\{\boldsymbol{\sigma}\}}\frac{d}{dt}\left[\mathcal{O}(\boldsymbol{\sigma}) p(\boldsymbol{\sigma};t)\right] &=: \frac{d}{dt} \langle \mathcal{O} \rangle_t \notag \\
        &= \sum_{\{\boldsymbol{\sigma}\}}\sum_{i} \left[\omega(\sigma^i) \mathcal{O}(\boldsymbol{\sigma})p(\boldsymbol{\sigma}^i;t) - \omega(\sigma_i)\mathcal{O}(\boldsymbol{\sigma}) p(\boldsymbol{\sigma};t)\right].
    \end{align}
    In order to perform the sum over all the configurations, we will separate this sums into two, so that
    \begin{align}
        \frac{d \langle \mathcal{O}\rangle_t}{dt} = \sum_{\{\boldsymbol{\sigma}\}}\sum_{i} \mathcal{O}(\boldsymbol{\sigma}) \omega(\sigma^i)p(\boldsymbol{\sigma}^i;t) - \sum_{\{\boldsymbol{\sigma}\}}\sum_{i}\mathcal{O}(\boldsymbol{\sigma}) \omega(\sigma_i) p(\boldsymbol{\sigma};t). \notag
    \end{align}
    The term on the right is just the expected value of $\mathcal{O}(\boldsymbol{\sigma}) \omega(\sigma_i)$. The term on the left is a little harder. However, since we are summing over all the possible configurations, we can make the change of variables $\boldsymbol{\sigma} \rightarrow \boldsymbol{\sigma} ^i$, so that $\boldsymbol{\sigma}^i \rightarrow (\boldsymbol{\sigma}^i)^i = \boldsymbol{\sigma}$, and, as a consequence,
    \begin{align}\label{eq:der_exp_O}
        \frac{d\langle \mathcal{O}\rangle_t}{dt} &= \sum_{\{\boldsymbol{\sigma}^i\}}\sum_{i}\mathcal{O}(\boldsymbol{\sigma}^i)\omega((\sigma^i)^i)p((\boldsymbol{\sigma}^i)^i;t) - \sum_{\{\boldsymbol{\sigma}\}}\sum_{i}\mathcal{O}(\boldsymbol{\sigma})\omega(\sigma_i)p(\boldsymbol{\sigma}\;t) \notag \\
        &= \sum_{i}\sum_{\{\boldsymbol{\sigma}^i\}} \mathcal{O}(\boldsymbol{\sigma}^i) \omega(\sigma_i)p(\boldsymbol{\sigma};t) - \sum_{i}\sum_{\{\boldsymbol{\sigma}\}} \mathcal{O}(\boldsymbol{\sigma})\omega(\sigma_i)p(\boldsymbol{\sigma};t) \notag \\
        &= \sum_{i} \langle \mathcal{O}(\boldsymbol{\sigma}^i) \omega(\sigma_i) \rangle_t - \sum_{i}\langle \mathcal{O}(\boldsymbol{\sigma}) \omega(\sigma_i)  \rangle_t = \boxed{\sum_{i}\langle [\mathcal{O}(\boldsymbol{\sigma}^i) - \mathcal{O}(\boldsymbol{\sigma})] \omega(\sigma_i) \rangle_t.}
    \end{align}
    If we now set $\mathcal{O}(\boldsymbol{\sigma}) = \sigma_k$ or $\mathcal{O}(\boldsymbol{\sigma}) = \sigma_k \sigma_{\ell}$, we can easily find the time evolution of the average spin and the correlation function.
    \begin{enumerate}
        \item For instance, by taking $\mathcal{O}(\boldsymbol{\sigma}) = \sigma_k$, we will have that $d\langle \mathcal{O}\rangle_t/dt = d\langle \sigma_k \rangle_t/dt$. Besides, since $\sigma^{i} := - \sigma_i$, $\mathcal{O}(\boldsymbol{\sigma}^i) - \mathcal{O}(\boldsymbol{\sigma}) = (\sigma^k - \sigma_k )\delta_{ik} =: -2\sigma_k \delta_{ik}$, as the difference $\sigma_k(\boldsymbol{\sigma}^i) - \sigma_k(\boldsymbol{\boldsymbol{\sigma}})$ will only be different than zero if the spin flipped is spin $k$ (note how we sum over every single flip of spin $i$). Thus, now equation Eq. (\ref{eq:der_exp_O}) reads,
        \begin{align}
            \frac{d \langle \sigma_k \rangle_t}{dt} = \sum_{i}\langle (-2\sigma_k)\delta_{ik}\omega(\sigma_i) \rangle_t = - 2\langle \sigma_k \omega(\sigma_k)\rangle_t.
        \end{align}
        \item Samewise, by taking $\mathcal{O} = \sigma_k \sigma_{\ell}$, we will have, $\mathcal{O}(\boldsymbol{\sigma}^i) - \mathcal{O}(\boldsymbol{\sigma}_i) =  -2\sigma_{k}\sigma_{\ell}(\delta_{ik}+\delta_{i\ell})$ since now the difference $\mathcal{O}(\boldsymbol{\sigma}^i) - \mathcal{O}(\boldsymbol{\sigma})$ will only be different than zero if either $i = k$ or $i = \ell$. As a consequence, we will have,
        \begin{align}
            \frac{d \langle \sigma_k \sigma_{\ell}\rangle_t}{dt} = \sum_{i}\langle -2\sigma_k\sigma_{\ell}(\delta_{ik} + \delta_{i\ell}) \omega(\sigma_i)\rangle_t = -2\langle \sigma_{k}\sigma_{\ell}[\omega(\sigma_k) + \omega(\sigma_{\ell})]\rangle_t.
        \end{align}
        \end{enumerate}

\section{Average of an Analytic Function of an Observable}\label{sec:mf-analytic-function}
Consider an analytic function of an observable, $\mathcal{O}(\boldsymbol{\sigma})$, $f(\mathcal{O}(\boldsymbol{\sigma}))$. For now, let us consider that the observable is nothing but the local field acting on spin $\sigma_i$, defined as $\mathcal{O}(\boldsymbol{\sigma}):= \beta h_i$, where $h_i$ is defined as in Eq. (\ref{eq:local_field}). Considering furthermore that $h_i^{\text{ext}} = 0$ and since $f$ is analytic,

\begin{equation}
    f(\beta h_i) = \sum_{n = 0}^{\infty} a_n (\beta h_i - (\beta h_i)_0)^n.
\end{equation}
Using the binomial theorem, the average value of $f(\beta h_i)$ thus can be written as,
\begin{align}\label{eq:avg-f}
    \langle f(\beta h_i) \rangle &= \sum_{n = 0}^{\infty}a_n \Big\langle (\beta h_i - (\beta h_i)_0)^n \Big \rangle = \sum_{n=0}^{\infty}a_n\sum_{k = 0}^{n} \binom{n}{k} \Big \langle (\beta h_i)^{n-k}\Big \rangle[-(\beta h_i)_0]^k.
\end{align}
Using the definition of $h_i$, as in Eq. (\ref{eq:local_field}), 
\begin{align}
    (\beta h_i)^{n-k} &= \left(\beta \sum_{j_1\in \langle i \rangle }J_{ij_1}\sigma_j\right)\left(\beta \sum_{j_2 \in \langle i \rangle}J_{ij_2}\sigma_{j_2}\right)\cdots\left(\beta \sum_{j_{n-k}\in \langle i \rangle} J_{ij_{n-k}} \sigma_{j_{n-k}}\right) \notag \\
    &= \beta^{n-k}\sum_{j_1,j_2,\dots,j_{n-k}\in \langle i \rangle}J_{ij_1}J_{ij_2}\dots J_{ij_{n-k}}\sigma_{j_1}\sigma_{j_2}\dots\sigma_{j_{n-k}},
\end{align}
such that the average value, 
\begin{align}
    \Big\langle (\beta h_{i})^{n-k} \Big\rangle = \beta^{n-k}\sum_{j_1,j_2,\dots,j_{n-k}\in \langle i \rangle}J_{ij_1}J_{ij_2}\dots J_{ij_{n-k}} \langle \sigma_{j_1}\sigma_{j_2}\dots\sigma_{j_{n-k}} \rangle,
\end{align}
along with the mean--field approximation $\langle \sigma_{j_1}\sigma_{j_2}\dots \sigma_{j_{n-k}}\rangle = \langle \sigma_{j_1}\rangle\langle\sigma_{j_2}\dots\sigma_{j_{n-k}}\rangle =  \dots = \langle \sigma_{j_1} \rangle \langle \sigma_{j_2}\rangle \dots \langle \sigma_{j_{n-k}}\rangle$, takes to
\begin{align}
    \Big\langle (\beta h_i)^{n-k} \Big\rangle &= \beta^{n-k}\sum_{j_1,j_2,\dots,j_{n-k}\in \langle i \rangle}J_{ij_1}J_{ij_2}\dots J_{ij_{n-k}} \langle \sigma_{j_1}\rangle\langle\sigma_{j_2}\rangle\dots\langle\sigma_{j_{n-k}} \rangle \notag \\
    &=\left(\beta \sum_{j_1 \in \langle i \rangle}J_{ij_1}\langle \sigma_{j_1}\rangle\right)\left(\beta \sum_{j_2 \in \langle i \rangle}J_{ij_2}\langle \sigma_{j_2}\rangle\right)\dots\left(\beta \sum_{j_{n-k} \in \langle i \rangle}J_{ij_{n-k}}\langle \sigma_{j_{n-k}}\rangle\right) \notag \\
    &=: (\beta \langle h_i \rangle)^{n-k}.
\end{align}
Plugging now that $\langle (\beta h_i)^{n-k} \rangle = (\beta \langle h_i \rangle)^{n-k}$ in Eq. (\ref{eq:avg-f}), 
\begin{align}
    \langle f(\beta h_i) \rangle = \sum_{n = 0}^{\infty} a_n \sum_{k = 0}^{n}\binom{n}{k}(\beta \langle h_i \rangle)^{n-k}[-(\beta h_i)_0]^{k} = \sum_{n=0}^{\infty}a_n (\beta \langle h_i \rangle - (\beta h_i)_0)^n =: f(\beta \langle h_i \rangle).
\end{align}
We thus have shown that in the mean--field approximation any analytic function $f(\beta h_i)$ verifies that $\langle f(\beta h_i) \rangle = f(\beta \langle h_i \rangle)$. \\

As done in the body of the article, this justifies using that $\langle \tanh \beta h_i \rangle = \tanh \beta \langle h_i \rangle$ since the hyperbolic tangent is analtyic. 

\section{Mean--Field Vision--Cone Model}\label{sec:mf-vc}

Using the definition for the Vision Cone (VC) model, $\rho_{ij} = \delta_{(\sigma_i,-)}\Theta(j<i) + \delta_{(\sigma_i,+)}\Theta(j>i)$ the dynamical equation for $m$ can be computed. First note how since $\sigma_i$ is an Ising variable, the Kronecker deltas in the definitions can be written as,
\begin{equation}
    \delta_{(\sigma_i,-)} = \frac{1 - \sigma_i}{2}, \quad \delta_{(\sigma_i,+)} = \frac{1+\sigma_i}{2}.
\end{equation}
We will then have that,
\begin{align}
    \langle \sigma_i \delta\omega_{i}\rangle & = -\frac{\beta}{2} \sech^2 \beta NJ m \sum_{j} \langle \rho_{ij}\sigma_j\rangle \notag \\
    &= - \frac{\beta}{2}\sech^2 (\beta NJm) \sum_{j}\langle [\delta_{(\sigma_i,-)}\Theta(j<i) + \delta_{(\sigma_i,+)} \Theta(j>i) ] \sigma_j\rangle  \notag \\
    & = -\frac{\beta}{4}\sech^2(\beta JNm)\left[ \sum_{j<i} \Big\langle \left(\frac{1-\sigma_i}{2}\right)\sigma_j \Big \rangle + \sum_{j>i} \Big\langle \left(\frac{1+\sigma_i}{2}\right)\sigma_j \Big \rangle\right]
\end{align}
We can now take the MF approximation $\langle \sigma_i \sigma_j\rangle = \langle \sigma_i \rangle \langle \sigma_j\rangle$ for $j\neq i$. By doing so, one obtains,
\begin{align}
    \langle \sigma_i \delta\omega_i\rangle = -\frac{\beta}{2}\sech^2(\beta NJ m)\left[ \sum_{j\neq i} \langle \sigma_j \rangle  + \langle \sigma_i \rangle \left(\sum_{j>i}\langle \sigma_j \rangle - \sum_{j\leq i}\langle \sigma_j \rangle \right)\right].
\end{align}
The first sum of the latter can be directly approximated by $Nm$. The two other sums, however, are not as trivial. As expected, spin $i$ divide a chain of spins into two, and so it appears to be the upper and lower bound of both. To perform this sums, we have to take the thermodynamic limit when $J := 1/N$ and $N\rightarrow \infty$. In the thermodynamic limit one can consider the two halves to be identical. 
Then, both sums will converge to the same object and we will be able to, in this case, identify every spin of both chains (separated by $i$) as identical, and of mean $m = \langle \sigma_j \rangle$ for all $j\neq i$. We will make the assumption that if $N$ is big enough, this approximation still holds, and then take the thermodynamic limit. In this situation, then
\begin{align}
    \langle \sigma_i \delta \omega_i \rangle &\approx - \frac{\beta}{2}\sech^2(\beta NJm) \left[Nm + \langle \sigma_i \rangle \left( \sum_{j>i}m  - \sum_{j<i}m\right)\right] \notag \\
    &=-\frac{\beta}{4} \sech^2(\beta NJm) [Nm + \langle \sigma_i \rangle m(N - (i+1) - (i-1))] \notag \\
    &=-\frac{\beta N m}{4}\left[1 + \langle \sigma_i \rangle\left(1 - 2\frac{i}{N}\right)\right]\sech^2(\beta Njm).
\end{align}
In here $i/N$ is the relative position of spin $\sigma_i$ on the chain. However, when making the MF approximation, the non-reciprocal coupling cannot have a spatial interpretation anymore. Taking into account that $i/N$ also represents the fraction of spins that are being overlooked by $\sigma_i$ when considering a fully connected model (since its relative position is $i/N$), we will define $f:=i/N$ the fraction of overlooked neighbours as a MF parameter, so that when we take the thermodynamic limit, we still have a well defined way of concieving a vision field. By doing so, again, we will have, then
\begin{align}
    \langle \sigma_i \delta \omega_i \rangle = -\frac{\beta Nm}{4}[1+\langle \sigma_i\rangle(1-2f)]\sech^2(\beta NJm).
\end{align}
We can hence add this one to the contribution of the symmetric (reciprocal) field in order to get the dynamic equation for $\langle \sigma_i \rangle$ as portrayed in equation Eq. (\ref{eq:dyn_si}), then divide by $N$ and sum over all spins to find the dynamic equation of $m$.

\section{Continuum--Approximation of the Vision--Cone model}\label{sec:app_e}

We take as starting point Eq. (\ref{eq:finite_d_mf_evo}) from the main text,
\begin{equation}
    \frac{\partial m_i}{\partial t} = -m_i + \tanh \beta\left( \sum_{j\in\langle i \rangle}(J +  \kappa \langle \rho_{ij} \rangle)m_j\right).
\end{equation}
Note how $i \in Z^d$ can be represented with position vector $\textbf{x}_i \in \mathbb{R}^d$ with cartesian base $\{\hat{e}_\alpha\}_{\alpha = 1,\dots,d }$. We can then make the continuum approximation considering that the spacing between spins is nothing but some characteristic length $a$ (for simplicity in the hypercubic base $|\hat{e}_{\alpha}| = a$, for any $\alpha=1,\dots,d$), that can be taken arbitrarily small, leading to
\begin{equation}
    \sum_{j\in\langle i \rangle} m_j \rightarrow 2dm(\textbf{x}_i) + a^2 \nabla^2 m(\textbf{x}_i).
\end{equation}

The contribution of $J_{ij}^a$ won't be as easy to compute and will depend on the model one uses. For a $f=1/2$ vision$-$cone interaction as defined in Eq. (\ref{eq:rhoVC}) in the main text, the term $\langle  \rho_{ij}(\boldsymbol{\sigma}) \rangle$ cannot be taken out of the sum, since depending on the state the contribution of the neighbours is different, accounting only for $j = i+\hat{e}_{x,y}$ or $j = i - \hat{e}_{x,y}$. Using that $\delta_{\sigma_i,\pm} = (1\pm \sigma_i)/2$, we have
\begin{align*}
    \kappa \sum_{j \in \langle i \rangle}\langle \rho_{ij}(\boldsymbol{\sigma}) \rangle m_j &= \kappa \sum_{j \in \langle i \rangle} \left(\frac{1+m_i}{2}\right)(\delta_{i+\hat{e}_x,j}+\delta_{i+\hat{e}_y,j})m_j + \kappa \sum_{j \in \langle i \rangle} \left(\frac{1-m_i}{2}\right)(\delta_{i-\hat{e}_x,j}+\delta_{i-\hat{e}_y,j})m_j \\
    & =\kappa \left(\frac{1+m_i}{2}\right)(m_{i+\hat{e}_x} + m_{i+\hat{e}_y}) + \kappa \left( \frac{1-m_i}{2}\right)(m_{i-\hat{e}_x} + m_{i-\hat{e}_y}) \\
    &= \kappa\left[\left(\frac{1+m_i}{2}\right)\left(\frac{m_{i+\hat{e}_x}-m_i}{a} a + \frac{m_{i+\hat{e}_y}-m_i}{a}a + 2m_i\right)\right]  \\
    &\;+\kappa \left[\left(\frac{1-m_i}{2}\right)\left(\frac{m_{i-\hat{e}_x}-m_i}{a} a + \frac{m_{i-\hat{e}_y}-m_i}{a}a + 2m_i\right)\right] ,
\end{align*}
where $a$ is, again, the spacing of the lattice. Identifying now the terms $(m_{i+\hat{e}_{x,y}}-m_i)/a \rightarrow \partial_{x,y} m_i$ and $(m_{i-\hat{e}_{x,y}} -m_i)/a \rightarrow -\partial_{x,y}m_i$ (where $i$ is can be directly represented by $\textbf{x}_i \in \mathbb{R}^2$), we can write the last the following way,
\begin{align*}
    \sum_{j\in\langle i \rangle}  \langle J_{ij}^a \rangle m_j = \kappa\left[2 m_i + \frac{1}{2}a m_i (\partial_x m_i + \partial_y m_i)\right].
\end{align*}
Identifying now the index $i$ with the vector $\textbf{x}_i$ in the base $\hat{e}_\alpha$ the continuum limit writes,
\begin{align}\label{eq:cont_sm_f_1/2}
    \sum_{j\in\langle i \rangle}  \langle J_{ij}^a \rangle m_j \rightarrow \kappa d m(\textbf{x}_i) + \frac{1}{2}\kappa a m(\textbf{x}_i)\left(\frac{\partial m(\textbf{x}_i)}{\partial x} + \frac{\partial m(\textbf{x}_i)}{\partial y}\right).
\end{align}
Note how we initially chose $\sigma_i = +1$ to interact with $i+\hat{e}_{x,y}$ and $\sigma_i = - 1$ to interact with $i-\hat{e}_{x,y}$ instead. That means that, if we set $J = 0$ the interaction is limited to the direction $(1,1) := \hat{e}_x + \hat{e}_y$ and the direction of interaction depends on the sign of $\sigma_i$ (positive is $\sigma_i = +1$ and negative otherwise). Note how this direction $\hat{e}_x + \hat{e}_y$ is the one that determines the sign of the derivatives in Eq. (\ref{eq:cont_sm_f_1/2}). We can then write,
\begin{align}
    m(\textbf{x}_i) \left(\frac{\partial m(\textbf{x}_i)}{\partial x} + \frac{\partial m(\textbf{x}_i)}{\partial y}\right) = (m(\textbf{x}_i) \textbf{v} \cdot \nabla)m(\textbf{x}_i),
\end{align}
where $\textbf{v} = (1,1)^T := \hat{e}_x + \hat{e}_y$. Doing the same thing we did for the previous one, we can introduce this continuum approximation in the corresponding term showing up in Eq. (\ref{eq:finite_d_mf_evo}) in the main text, and Taylor expand the result along small $m$ and its possible gradients, keeping them at first order, we obtain,
\begin{align}
    \frac{\partial m(\textbf{x},t)}{\partial t} = &-(1-2d\beta J - d\beta \kappa)m(\textbf{x},t) - \frac{1}{3}\beta^3(d(2J + \Delta))^3 m(\textbf{x},t)^3 \notag \\
    &+ a^2 \beta J \nabla^2 m(\textbf{x},t) + \frac{1}{2}a \beta \kappa (m\textbf{v}\cdot \nabla)m(\textbf{x},t).
\end{align}
Eq. (\ref{eq:continuum_vc_finite_d}) portrays the competition between diffusion and advection, the diffusion comes from the usual Ising (reciprocal) coupling between first neighbours in an homogenous manner, and the second one, coupled to $\kappa$, which shows advection in a direction that explicitly depends on the state of local magnetization.\\

The continuum limit might be taken for $f = 3/4$ as well trivially, taking into account now that second order derivatives (in discrete form) will appear.

\section{Coarsening of Non--Conserved Fields}\label{sec:app_f}
Consider a non--conserved scalar field $\phi(\textbf{x},t)$ with (deterministic, in the absence of thermal noise) time evolution given by \cite{bray2002theory}

\begin{align}
    \frac{\partial \phi (\textbf{x},t)}{\partial t} = -\frac{\delta F[\phi]}{\delta \phi(\textbf{x},t)}, \quad F[\phi] = \int d^d{\textbf{x}} \left(\frac{1}{2}(\nabla \phi)^2 + V(\phi)\right), 
\end{align}
where $V(\phi) = (1-\phi^2)^2$ is a double well potential with $V(\pm 1) = 0$. The contribution of the term $(\nabla \phi)^2$ in $F[\phi]$ corresponds to a cost of energy to sustain domain walls (interfaces) between phases $\phi = +1$ and $\phi=-1$. 

\begin{figure}[h!]
    \centering
    \includegraphics[width=0.9\linewidth]{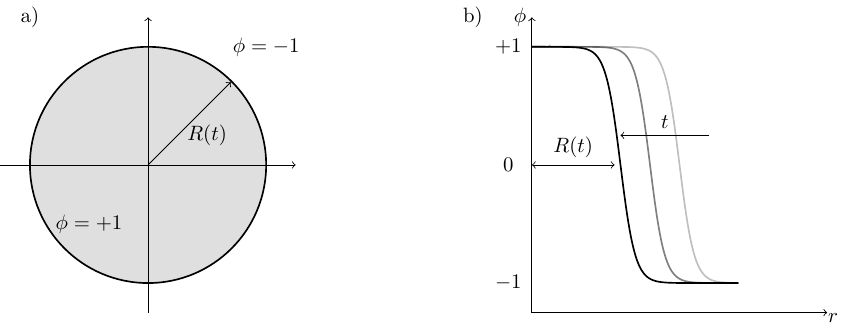}
    \caption{(a) Sketch of a drop of evolving radius $R(t)$ in terms of a field $\phi$ which takes values $\phi = +1$ inside, $r< R(t)$ and $\phi = - 1$ outside, $r>R(t)$. The drop shrinks due to the presence of the sea $\phi = -1$ surrounding it, making $R(t)$ to decrease with time; (b) dependence of the field with the radius. Note how the interface between $\phi = + 1$ and $\phi = -1$ is placed at $R(t)$ and is very sharp around the interface.}
    \label{fig:sketch_drop}
\end{figure}

Consider now a configuration in the shape of a \textit{drop} of $\phi = +1$ immersed in a sea of $\phi = -1$ as shown in Fig. (\ref{fig:sketch_drop}). The spherical symmetry provided by the configuration allows the ansatz $ \phi(\textbf{x}, t) =: \phi(r,t)$ where now $r$ is the radial coordinate. Exploiting radial symmetry, the evolution equation becomes \cite{bray2002theory},
\begin{align}
    \frac{\partial \phi(r,t)}{\partial t} = \frac{\partial^2 \phi(r,t)}{\partial r^2} + \frac{d-1}{r}\frac{\partial \phi(r,t)}{\partial r} -V'(\phi).
\end{align}
Given that the radius of the drop $R(t)$ is much bigger than the interface width, $\xi$, 
 and considering that the time dependence of $\phi$ is simply absorbed by the existence of an interface at $r = R(t)$, we assume $\phi(r,t)$ to have the shape $\phi(r,t) = g(r- R(t))$, where $g(z)$ sharply goes from $+1$ to $-1$ when crossing $z = 0$ (the  interface width is negligible compared to $R$). The partial derivatives can now be computed to give $\partial_t \phi(r,t) = -g' (dR/{dt})$, $\partial_r \phi = g'$ and $\partial_r^2  \phi(r,t) = g''$. We thus have,
\begin{equation}\label{eq:coarsening_g''}
    g'' + \left[\frac{d-1}{r} + \frac{dR}{dt}\right]g' - V'(\phi) = 0.
\end{equation}
Multiplying by $g'$ and integrating Eq. (\ref{eq:coarsening_g''}) across the interface, $z = 0$,
\begin{align}
    \lim_{\epsilon \rightarrow 0} \int_{-\epsilon}^{\epsilon}g''(z)g'(z)\;dz + \int_{-\epsilon}^{\epsilon} \left[\frac{d-1}{r} + \frac{dR}{dt}\right](g')^2(z)\; dz - \int_{-\epsilon}^{\epsilon} V'(g(z))g'(z)\; dz = 0. 
\end{align}
We then use that $g'(z) \approx 0$ for any $|z| \gg \xi$, $r = z + R(t) \approx R(t)$, $g(z) \approx +1$ when $z <0$, $g(z) \approx - 1$ for $z>0$, and that $V(\pm 1) = 0$, to derive
\begin{align}\label{eq:evo_radius}
    \frac{d-1}{R} + \frac{dR}{dt} = 0. 
\end{align}
A solution of this equation is $R(t)^2 = R_0^2 - 2(d-1)t$, where $R_0 =: R(0)$ is the initial size of the drop. This is only valid as long as $R(t)$ remains bigger than the size of the interface $\xi$, however, it also provides a good approximation of the \textit{shrinking time} that it takes for a drop to completely disappear, $R(\tau_s) = 0 = R_0^2-2(d-1)\tau_s$, or $\tau_s \sim R_0^2$. \\

We can also use the solution of Eq. (\ref{eq:evo_radius}) to understand how the magnetization of a lattice of Ising spins behaves in time when it evolves from an initially ordered system consisting of a drop of spins pointing up $\sigma_i = +1$ with initial radius $R_0$ lays in a sea of spins pointing down, $\sigma_i = -1$; for a $d = 2$ lattice of size $L$, the global magnetization will be nothing but the difference of the fraction of area occupied by spins pointing in different direction. The area occupied by $+1$ spins is $\pi R(t)^2$ while the one occupied by spins pointing down is $L^2 - \pi R(t)^2$. Thus, the global magnetization becomes $m(t) = \pi (R(t)/L)^2 - (L^2 - \pi R(t)^2)/L^2 = 2 \pi (R(t)/L) ^2 - 1$. Since, when $R(t) \gg \xi, R^2(t) = R_0 ^2 - 2(d-1)t$, we find
\begin{align}
    m(t) = \left[2\pi \left(\frac{R_0}{L}\right)^2 - 1 \right] - \frac{4\pi (d-1)}{L^2}t =: m_0 - At,
\end{align}
and so, $m(t)$ evolves linearly with $t$.




\end{appendices}

\newpage
\bibliography{sn-bibliography} 


\begin{thebibliography}{54}
\ifx \bisbn   \undefined \def \bisbn  #1{ISBN #1}\fi
\ifx \binits  \undefined \def \binits#1{#1}\fi
\ifx \bauthor  \undefined \def \bauthor#1{#1}\fi
\ifx \batitle  \undefined \def \batitle#1{#1}\fi
\ifx \bjtitle  \undefined \def \bjtitle#1{#1}\fi
\ifx \bvolume  \undefined \def \bvolume#1{\textbf{#1}}\fi
\ifx \byear  \undefined \def \byear#1{#1}\fi
\ifx \bissue  \undefined \def \bissue#1{#1}\fi
\ifx \bfpage  \undefined \def \bfpage#1{#1}\fi
\ifx \blpage  \undefined \def \blpage #1{#1}\fi
\ifx \burl  \undefined \def \burl#1{\textsf{#1}}\fi
\ifx \doiurl  \undefined \def \doiurl#1{\url{https://doi.org/#1}}\fi
\ifx \betal  \undefined \def \betal{\textit{et al.}}\fi
\ifx \binstitute  \undefined \def \binstitute#1{#1}\fi
\ifx \binstitutionaled  \undefined \def \binstitutionaled#1{#1}\fi
\ifx \bctitle  \undefined \def \bctitle#1{#1}\fi
\ifx \beditor  \undefined \def \beditor#1{#1}\fi
\ifx \bpublisher  \undefined \def \bpublisher#1{#1}\fi
\ifx \bbtitle  \undefined \def \bbtitle#1{#1}\fi
\ifx \bedition  \undefined \def \bedition#1{#1}\fi
\ifx \bseriesno  \undefined \def \bseriesno#1{#1}\fi
\ifx \blocation  \undefined \def \blocation#1{#1}\fi
\ifx \bsertitle  \undefined \def \bsertitle#1{#1}\fi
\ifx \bsnm \undefined \def \bsnm#1{#1}\fi
\ifx \bsuffix \undefined \def \bsuffix#1{#1}\fi
\ifx \bparticle \undefined \def \bparticle#1{#1}\fi
\ifx \barticle \undefined \def \barticle#1{#1}\fi
\bibcommenthead
\ifx \bconfdate \undefined \def \bconfdate #1{#1}\fi
\ifx \botherref \undefined \def \botherref #1{#1}\fi
\ifx \url \undefined \def \url#1{\textsf{#1}}\fi
\ifx \bchapter \undefined \def \bchapter#1{#1}\fi
\ifx \bbook \undefined \def \bbook#1{#1}\fi
\ifx \bcomment \undefined \def \bcomment#1{#1}\fi
\ifx \oauthor \undefined \def \oauthor#1{#1}\fi
\ifx \citeauthoryear \undefined \def \citeauthoryear#1{#1}\fi
\ifx \endbibitem  \undefined \def \endbibitem {}\fi
\ifx \bconflocation  \undefined \def \bconflocation#1{#1}\fi
\ifx \arxivurl  \undefined \def \arxivurl#1{\textsf{#1}}\fi
\csname PreBibitemsHook\endcsname

\bibitem[\protect\citeauthoryear{Klapp}{2023}]{sabine_living}
\begin{barticle}
\bauthor{\bsnm{Klapp}, \binits{S.H.L.}}:
\batitle{Non-reciprocal interaction for living matter}.
\bjtitle{Nat. Nanotechnol.}
\bvolume{18},
\bfpage{8}--\blpage{9}
(\byear{2023})
\end{barticle}
\endbibitem

\bibitem[\protect\citeauthoryear{Castellano
  et~al.}{2009}]{castellano2009statistical}
\begin{barticle}
\bauthor{\bsnm{Castellano}, \binits{C.}},
\bauthor{\bsnm{Fortunato}, \binits{S.}},
\bauthor{\bsnm{Loreto}, \binits{V.}}:
\batitle{Statistical physics of social dynamics}.
\bjtitle{Reviews of modern physics}
\bvolume{81}(\bissue{2}),
\bfpage{591}--\blpage{646}
(\byear{2009})
\end{barticle}
\endbibitem

\bibitem[\protect\citeauthoryear{Agudo-Canalejo and
  Golestanian}{2019}]{agudo-canalejo}
\begin{barticle}
\bauthor{\bsnm{Agudo-Canalejo}, \binits{J.}},
\bauthor{\bsnm{Golestanian}, \binits{R.}}:
\batitle{Active phase separation in mixtures of chemically interacting
  particles}.
\bjtitle{Phys. Rev. Lett.}
\bvolume{123},
\bfpage{018101}
(\byear{2019})
\doiurl{10.1103/PhysRevLett.123.018101}
\end{barticle}
\endbibitem

\bibitem[\protect\citeauthoryear{Saha et~al.}{2020}]{saha2020scalar}
\begin{barticle}
\bauthor{\bsnm{Saha}, \binits{S.}},
\bauthor{\bsnm{Agudo-Canalejo}, \binits{J.}},
\bauthor{\bsnm{Golestanian}, \binits{R.}}:
\batitle{Scalar active mixtures: The nonreciprocal cahn-hilliard model}.
\bjtitle{Physical Review X}
\bvolume{10}(\bissue{4}),
\bfpage{041009}
(\byear{2020})
\end{barticle}
\endbibitem

\bibitem[\protect\citeauthoryear{Schmidt et~al.}{2019}]{schmidt2019light}
\begin{botherref}
\oauthor{\bsnm{Schmidt}, \binits{F.}},
\oauthor{\bsnm{Liebchen}, \binits{B.}},
\oauthor{\bsnm{L{\"o}wen}, \binits{H.}},
\oauthor{\bsnm{Volpe}, \binits{G.}}:
Light-controlled assembly of active colloidal molecules.
The Journal of chemical physics
\textbf{150}(9)
(2019)
\end{botherref}
\endbibitem

\bibitem[\protect\citeauthoryear{Dinelli
  et~al.}{2023}]{dinelli2023nonreciprocal}
\begin{barticle}
\bauthor{\bsnm{Dinelli}, \binits{A.}},
\bauthor{\bsnm{O’Byrne}, \binits{J.}},
\bauthor{\bsnm{Curatolo}, \binits{A.}},
\bauthor{\bsnm{Zhao}, \binits{Y.}},
\bauthor{\bsnm{Sollich}, \binits{P.}},
\bauthor{\bsnm{Tailleur}, \binits{J.}}:
\batitle{Non-reciprocity across scales in active mixtures}.
\bjtitle{Nature Communications}
\bvolume{14}(\bissue{1}),
\bfpage{7035}
(\byear{2023})
\end{barticle}
\endbibitem

\bibitem[\protect\citeauthoryear{Ros et~al.}{2023}]{lotka-volterra}
\begin{barticle}
\bauthor{\bsnm{Ros}, \binits{V.}},
\bauthor{\bsnm{Roy}, \binits{F.}},
\bauthor{\bsnm{Biroli}, \binits{G.}},
\bauthor{\bsnm{Bunin}, \binits{G.}},
\bauthor{\bsnm{Turner}, \binits{A.M.}}:
\batitle{Generalized lotka-volterra equations with random, nonreciprocal
  interactions: The typical number of equilibria}.
\bjtitle{Phys. Rev. Lett.}
\bvolume{130},
\bfpage{257401}
(\byear{2023})
\doiurl{10.1103/PhysRevLett.130.257401}
\end{barticle}
\endbibitem

\bibitem[\protect\citeauthoryear{Blumenthal et~al.}{2024}]{chaos_eco}
\begin{barticle}
\bauthor{\bsnm{Blumenthal}, \binits{E.}},
\bauthor{\bsnm{Rocks}, \binits{J.W.}},
\bauthor{\bsnm{Mehta}, \binits{P.}}:
\batitle{Phase transition to chaos in complex ecosystems with nonreciprocal
  species-resource interactions}.
\bjtitle{Phys. Rev. Lett.}
\bvolume{132},
\bfpage{127401}
(\byear{2024})
\doiurl{10.1103/PhysRevLett.132.127401}
\end{barticle}
\endbibitem

\bibitem[\protect\citeauthoryear{Brandenbourger
  et~al.}{2019}]{brandenbourger2019non}
\begin{barticle}
\bauthor{\bsnm{Brandenbourger}, \binits{M.}},
\bauthor{\bsnm{Locsin}, \binits{X.}},
\bauthor{\bsnm{Lerner}, \binits{E.}},
\bauthor{\bsnm{Coulais}, \binits{C.}}:
\batitle{Non-reciprocal robotic metamaterials}.
\bjtitle{Nature communications}
\bvolume{10}(\bissue{1}),
\bfpage{4608}
(\byear{2019})
\end{barticle}
\endbibitem

\bibitem[\protect\citeauthoryear{March et~al.}{2023}]{march2023honeybee}
\begin{botherref}
\oauthor{\bsnm{March}, \binits{D.}},
\oauthor{\bsnm{M{\'u}gica}, \binits{J.}},
\oauthor{\bsnm{Ferrero}, \binits{E.E.}},
\oauthor{\bsnm{Miguel}, \binits{M.C.}}:
Honeybee-like collective decision making in a kilobot swarm.
arXiv preprint arXiv:2310.15592
(2023)
\end{botherref}
\endbibitem

\bibitem[\protect\citeauthoryear{Xu et~al.}{2023}]{non-hermitian}
\begin{barticle}
\bauthor{\bsnm{Xu}, \binits{G.}},
\bauthor{\bsnm{Zhou}, \binits{X.}},
\bauthor{\bsnm{Li}, \binits{Y.}},
\bauthor{\bsnm{Cao}, \binits{Q.}},
\bauthor{\bsnm{Chen}, \binits{W.}},
\bauthor{\bsnm{Xiao}, \binits{Y.}},
\bauthor{\bsnm{Yang}, \binits{L.}},
\bauthor{\bsnm{Qiu}, \binits{C.-W.}}:
\batitle{Non-hermitian chiral heat transport}.
\bjtitle{Phys. Rev. Lett.}
\bvolume{130},
\bfpage{266303}
(\byear{2023})
\doiurl{10.1103/PhysRevLett.130.266303}
\end{barticle}
\endbibitem

\bibitem[\protect\citeauthoryear{You et~al.}{2020}]{traveling_Marchetti}
\begin{barticle}
\bauthor{\bsnm{You}, \binits{Z.}},
\bauthor{\bsnm{Baskaran}, \binits{A.}},
\bauthor{\bsnm{Marchetti}, \binits{M.C.}}:
\batitle{Nonreciprocity as a generic route to traveling states}.
\bjtitle{Proceedings of the National Academy of Sciences}
\bvolume{117}(\bissue{33}),
\bfpage{19767}--\blpage{19772}
(\byear{2020})
\end{barticle}
\endbibitem

\bibitem[\protect\citeauthoryear{Loos and
  Klapp}{2020}]{loos_irreversibility_2020}
\begin{botherref}
\oauthor{\bsnm{Loos}, \binits{S.A.M.}},
\oauthor{\bsnm{Klapp}, \binits{S.H.L.}}:
Irreversibility, heat and information flows induced by non-reciprocal
  interactions.
NJP
\textbf{22}(123051)
(2020)
\end{botherref}
\endbibitem

\bibitem[\protect\citeauthoryear{Fruchart et~al.}{2021}]{Fruchart2021}
\begin{barticle}
\bauthor{\bsnm{Fruchart}, \binits{M.}},
\bauthor{\bsnm{Hanai}, \binits{R.}},
\bauthor{\bsnm{Littlewood}, \binits{P.B.}},
\bauthor{\bsnm{Vitelli}, \binits{V.}}:
\batitle{Non-reciprocal phase transitions}.
\bjtitle{Nature}
\bvolume{592},
\bfpage{363}--\blpage{369}
(\byear{2021})
\end{barticle}
\endbibitem

\bibitem[\protect\citeauthoryear{Zhang and
  Garcia-Millan}{2023}]{zhang2023entropy}
\begin{barticle}
\bauthor{\bsnm{Zhang}, \binits{Z.}},
\bauthor{\bsnm{Garcia-Millan}, \binits{R.}}:
\batitle{Entropy production of nonreciprocal interactions}.
\bjtitle{Physical Review Research}
\bvolume{5}(\bissue{2}),
\bfpage{022033}
(\byear{2023})
\end{barticle}
\endbibitem

\bibitem[\protect\citeauthoryear{Suchanek et~al.}{2023a}]{suchanek_23}
\begin{barticle}
\bauthor{\bsnm{Suchanek}, \binits{T.}},
\bauthor{\bsnm{Kroy}, \binits{K.}},
\bauthor{\bsnm{Loos}, \binits{S.A.M.}}:
\batitle{Entropy production in the nonreciprocal cahn-hilliard model}.
\bjtitle{Phys. Rev. E}
\bvolume{108},
\bfpage{064610}
(\byear{2023})
\doiurl{10.1103/PhysRevE.108.064610}
\end{barticle}
\endbibitem

\bibitem[\protect\citeauthoryear{Suchanek
  et~al.}{2023b}]{suchanek2023irreversible}
\begin{barticle}
\bauthor{\bsnm{Suchanek}, \binits{T.}},
\bauthor{\bsnm{Kroy}, \binits{K.}},
\bauthor{\bsnm{Loos}, \binits{S.A.}}:
\batitle{Irreversible mesoscale fluctuations herald the emergence of dynamical
  phases}.
\bjtitle{Physical Review Letters}
\bvolume{131}(\bissue{25}),
\bfpage{258302}
(\byear{2023})
\end{barticle}
\endbibitem

\bibitem[\protect\citeauthoryear{Osat and Golestanian}{2023}]{osat2023non}
\begin{barticle}
\bauthor{\bsnm{Osat}, \binits{S.}},
\bauthor{\bsnm{Golestanian}, \binits{R.}}:
\batitle{Non-reciprocal multifarious self-organization}.
\bjtitle{Nature Nanotechnology}
\bvolume{18}(\bissue{1}),
\bfpage{79}--\blpage{85}
(\byear{2023})
\end{barticle}
\endbibitem

\bibitem[\protect\citeauthoryear{Theveneau et~al.}{2013}]{theveneau2013chase}
\begin{barticle}
\bauthor{\bsnm{Theveneau}, \binits{E.}},
\bauthor{\bsnm{Steventon}, \binits{B.}},
\bauthor{\bsnm{Scarpa}, \binits{E.}},
\bauthor{\bsnm{Garcia}, \binits{S.}},
\bauthor{\bsnm{Trepat}, \binits{X.}},
\bauthor{\bsnm{Streit}, \binits{A.}},
\bauthor{\bsnm{Mayor}, \binits{R.}}:
\batitle{Chase-and-run between adjacent cell populations promotes directional
  collective migration}.
\bjtitle{Nature cell biology}
\bvolume{15}(\bissue{7}),
\bfpage{763}--\blpage{772}
(\byear{2013})
\end{barticle}
\endbibitem

\bibitem[\protect\citeauthoryear{Ivlev et~al.}{2015}]{ivlev2015statistical}
\begin{barticle}
\bauthor{\bsnm{Ivlev}, \binits{A.V.}},
\bauthor{\bsnm{Bartnick}, \binits{J.}},
\bauthor{\bsnm{Heinen}, \binits{M.}},
\bauthor{\bsnm{Du}, \binits{C.-R.}},
\bauthor{\bsnm{Nosenko}, \binits{V.}},
\bauthor{\bsnm{L{\"o}wen}, \binits{H.}}:
\batitle{Statistical mechanics where newton’s third law is broken}.
\bjtitle{Physical Review X}
\bvolume{5}(\bissue{1}),
\bfpage{011035}
(\byear{2015})
\end{barticle}
\endbibitem

\bibitem[\protect\citeauthoryear{Guislain and
  Bertin}{2024a}]{guislain_mean_field}
\begin{barticle}
\bauthor{\bsnm{Guislain}, \binits{L.}},
\bauthor{\bsnm{Bertin}, \binits{E.}}:
\batitle{Discontinuous phase transition from ferromagnetic to oscillating
  states in a nonequilibrium mean-field spin model}.
\bjtitle{Phys. Rev. E}
\bvolume{109},
\bfpage{034131}
(\byear{2024})
\doiurl{10.1103/PhysRevE.109.034131}
\end{barticle}
\endbibitem

\bibitem[\protect\citeauthoryear{Guislain and
  Bertin}{2024b}]{Guislain_2024_spin}
\begin{barticle}
\bauthor{\bsnm{Guislain}, \binits{L.}},
\bauthor{\bsnm{Bertin}, \binits{E.}}:
\batitle{Collective oscillations in a three-dimensional spin model with
  non-reciprocal interactions}.
\bjtitle{Journal of Statistical Mechanics: Theory and Experiment}
\bvolume{2024}(\bissue{9}),
\bfpage{093210}
(\byear{2024})
\doiurl{10.1088/1742-5468/ad72dc}
\end{barticle}
\endbibitem

\bibitem[\protect\citeauthoryear{Avni et~al.}{2023}]{avni2023non}
\begin{botherref}
\oauthor{\bsnm{Avni}, \binits{Y.}},
\oauthor{\bsnm{Fruchart}, \binits{M.}},
\oauthor{\bsnm{Martin}, \binits{D.}},
\oauthor{\bsnm{Seara}, \binits{D.}},
\oauthor{\bsnm{Vitelli}, \binits{V.}}:
The non-reciprocal ising model.
arXiv preprint arXiv:2311.05471
(2023)
\end{botherref}
\endbibitem

\bibitem[\protect\citeauthoryear{Couzin et~al.}{2003}]{couzin2003self}
\begin{barticle}
\bauthor{\bsnm{Couzin}, \binits{I.D.}},
\bauthor{\bsnm{Krause}, \binits{J.}}, \betal:
\batitle{Self-organization and collective behavior in vertebrates}.
\bjtitle{Advances in the Study of Behavior}
\bvolume{32}(\bissue{1}),
\bfpage{10}--\blpage{1016}
(\byear{2003})
\end{barticle}
\endbibitem

\bibitem[\protect\citeauthoryear{Peruani and
  Sibona}{2008}]{peruani2008dynamics}
\begin{barticle}
\bauthor{\bsnm{Peruani}, \binits{F.}},
\bauthor{\bsnm{Sibona}, \binits{G.J.}}:
\batitle{Dynamics and steady states in excitable mobile agent systems}.
\bjtitle{Physical review letters}
\bvolume{100}(\bissue{16}),
\bfpage{168103}
(\byear{2008})
\end{barticle}
\endbibitem

\bibitem[\protect\citeauthoryear{Bastien and
  Romanczuk}{2020}]{bastien2020model}
\begin{barticle}
\bauthor{\bsnm{Bastien}, \binits{R.}},
\bauthor{\bsnm{Romanczuk}, \binits{P.}}:
\batitle{A model of collective behavior based purely on vision}.
\bjtitle{Science advances}
\bvolume{6}(\bissue{6}),
\bfpage{0792}
(\byear{2020})
\end{barticle}
\endbibitem

\bibitem[\protect\citeauthoryear{Levis et~al.}{2020}]{levis2020flocking}
\begin{barticle}
\bauthor{\bsnm{Levis}, \binits{D.}},
\bauthor{\bsnm{Diaz-Guilera}, \binits{A.}},
\bauthor{\bsnm{Pagonabarraga}, \binits{I.}},
\bauthor{\bsnm{Starnini}, \binits{M.}}:
\batitle{Flocking-enhanced social contagion}.
\bjtitle{Physical Review Research}
\bvolume{2}(\bissue{3}),
\bfpage{032056}
(\byear{2020})
\end{barticle}
\endbibitem

\bibitem[\protect\citeauthoryear{Rodr{\'\i}guez
  et~al.}{2022}]{rodriguez2022epidemic}
\begin{barticle}
\bauthor{\bsnm{Rodr{\'\i}guez}, \binits{J.P.}},
\bauthor{\bsnm{Paoluzzi}, \binits{M.}},
\bauthor{\bsnm{Levis}, \binits{D.}},
\bauthor{\bsnm{Starnini}, \binits{M.}}:
\batitle{Epidemic processes on self-propelled particles: Continuum and
  agent-based modeling}.
\bjtitle{Physical Review Research}
\bvolume{4}(\bissue{4}),
\bfpage{043160}
(\byear{2022})
\end{barticle}
\endbibitem

\bibitem[\protect\citeauthoryear{G{\'o}mez-Nava
  et~al.}{2022}]{gomez2022intermittent}
\begin{barticle}
\bauthor{\bsnm{G{\'o}mez-Nava}, \binits{L.}},
\bauthor{\bsnm{Bon}, \binits{R.}},
\bauthor{\bsnm{Peruani}, \binits{F.}}:
\batitle{Intermittent collective motion in sheep results from alternating the
  role of leader and follower}.
\bjtitle{Nature Physics}
\bvolume{18}(\bissue{12}),
\bfpage{1494}--\blpage{1501}
(\byear{2022})
\end{barticle}
\endbibitem

\bibitem[\protect\citeauthoryear{Seara et~al.}{2023}]{seara2023}
\begin{barticle}
\bauthor{\bsnm{Seara}, \binits{D.S.}},
\bauthor{\bsnm{Piya}, \binits{A.}},
\bauthor{\bsnm{Tabatabai}, \binits{A.P.}}:
\batitle{Non-reciprocal interactions spatially propagate fluctuations in a 2d
  ising model}.
\bjtitle{Journal of Statistical Mechanics: Theory and Experiment}
\bvolume{2023}(\bissue{4}),
\bfpage{043209}
(\byear{2023})
\doiurl{10.1088/1742-5468/accce7}
\end{barticle}
\endbibitem

\bibitem[\protect\citeauthoryear{Rajeev and Kumar}{2024}]{rajeev2024ising}
\begin{botherref}
\oauthor{\bsnm{Rajeev}, \binits{A.K.}},
\oauthor{\bsnm{Kumar}, \binits{A.}}:
Ising model with non-reciprocal interactions.
arXiv preprint arXiv:2403.06875
(2024)
\end{botherref}
\endbibitem

\bibitem[\protect\citeauthoryear{Di~Carlo}{2024}]{di2024off}
\begin{botherref}
\oauthor{\bsnm{Di~Carlo}, \binits{L.}}:
The off-equilibrium kinetic ising model: The metric case.
arXiv preprint arXiv:2408.11690
(2024)
\end{botherref}
\endbibitem

\bibitem[\protect\citeauthoryear{Crisanti and
  Sompolinsky}{1988}]{CrisantiIsing}
\begin{barticle}
\bauthor{\bsnm{Crisanti}, \binits{A.}},
\bauthor{\bsnm{Sompolinsky}, \binits{H.}}:
\batitle{Dynamics of spin systems with randomly asymmetric bonds: Ising spins
  and glauber dynamics}.
\bjtitle{Phys. Rev. A}
\bvolume{37},
\bfpage{4865}--\blpage{4874}
(\byear{1988})
\doiurl{10.1103/PhysRevA.37.4865}
\end{barticle}
\endbibitem

\bibitem[\protect\citeauthoryear{Crisanti and
  Sompolinsky}{1987}]{CrisantiSpherical}
\begin{barticle}
\bauthor{\bsnm{Crisanti}, \binits{A.}},
\bauthor{\bsnm{Sompolinsky}, \binits{H.}}:
\batitle{Dynamics of spin systems with randomly asymmetric bonds: Langevin
  dynamics and a spherical model}.
\bjtitle{Phys. Rev. A}
\bvolume{36},
\bfpage{4922}--\blpage{4939}
(\byear{1987})
\doiurl{10.1103/PhysRevA.36.4922}
\end{barticle}
\endbibitem

\bibitem[\protect\citeauthoryear{Parisi}{1986}]{parisi1986asymmetric}
\begin{barticle}
\bauthor{\bsnm{Parisi}, \binits{G.}}:
\batitle{Asymmetric neural networks and the process of learning}.
\bjtitle{Journal of Physics A: Mathematical and General}
\bvolume{19}(\bissue{11}),
\bfpage{675}
(\byear{1986})
\end{barticle}
\endbibitem

\bibitem[\protect\citeauthoryear{Guislain and
  Bertin}{2024a}]{Guislain_disordered}
\begin{barticle}
\bauthor{\bsnm{Guislain}, \binits{L.}},
\bauthor{\bsnm{Bertin}, \binits{E.}}:
\batitle{Hidden collective oscillations in a disordered mean-field spin model
  with non-reciprocal interactions}.
\bjtitle{Journal of Physics A: Mathematical and Theoretical}
\bvolume{57}(\bissue{37}),
\bfpage{375001}
(\byear{2024})
\doiurl{10.1088/1751-8121/ad6ab4}
\end{barticle}
\endbibitem

\bibitem[\protect\citeauthoryear{Guislain and Bertin}{2024b}]{guislain2024far}
\begin{botherref}
\oauthor{\bsnm{Guislain}, \binits{L.}},
\oauthor{\bsnm{Bertin}, \binits{E.}}:
Far-from-equilibrium complex landscapes.
arXiv preprint arXiv:2405.08452
(2024)
\end{botherref}
\endbibitem

\bibitem[\protect\citeauthoryear{Hanai}{2024}]{hanai2024nonreciprocal}
\begin{barticle}
\bauthor{\bsnm{Hanai}, \binits{R.}}:
\batitle{Nonreciprocal frustration: Time crystalline order-by-disorder
  phenomenon and a spin-glass-like state}.
\bjtitle{Physical Review X}
\bvolume{14}(\bissue{1}),
\bfpage{011029}
(\byear{2024})
\end{barticle}
\endbibitem

\bibitem[\protect\citeauthoryear{Lorenzana et~al.}{2024}]{lorenzana2024non}
\begin{botherref}
\oauthor{\bsnm{Lorenzana}, \binits{G.G.}},
\oauthor{\bsnm{Altieri}, \binits{A.}},
\oauthor{\bsnm{Biroli}, \binits{G.}},
\oauthor{\bsnm{Fruchart}, \binits{M.}},
\oauthor{\bsnm{Vitelli}, \binits{V.}}:
Non-reciprocal spin-glass transition and aging.
arXiv preprint arXiv:2408.17360
(2024)
\end{botherref}
\endbibitem

\bibitem[\protect\citeauthoryear{Durve et~al.}{2018}]{durve2018active}
\begin{barticle}
\bauthor{\bsnm{Durve}, \binits{M.}},
\bauthor{\bsnm{Saha}, \binits{A.}},
\bauthor{\bsnm{Sayeed}, \binits{A.}}:
\batitle{Active particle condensation by non-reciprocal and time-delayed
  interactions}.
\bjtitle{The European Physical Journal E}
\bvolume{41},
\bfpage{1}--\blpage{9}
(\byear{2018})
\end{barticle}
\endbibitem

\bibitem[\protect\citeauthoryear{Loos et~al.}{2023}]{loos2023long}
\begin{barticle}
\bauthor{\bsnm{Loos}, \binits{S.A.}},
\bauthor{\bsnm{Klapp}, \binits{S.H.}},
\bauthor{\bsnm{Martynec}, \binits{T.}}
\bjtitle{Phys. Rev. Lett.}
\bvolume{130}(\bissue{19}),
\bfpage{198301}
(\byear{2023})
\end{barticle}
\endbibitem

\bibitem[\protect\citeauthoryear{Rouzaire et~al.}{2024}]{rouzaire2024non}
\begin{botherref}
\oauthor{\bsnm{Rouzaire}, \binits{Y.}},
\oauthor{\bsnm{Pagonabarraga}, \binits{I.}},
\oauthor{\bsnm{Pearce}, \binits{D.}},
\oauthor{\bsnm{Levis}, \binits{D.}}:
Non-reciprocal interactions reshape topological defect annihilation.
arXiv preprint arXiv:2401.12637
(2024)
\end{botherref}
\endbibitem

\bibitem[\protect\citeauthoryear{Huang et~al.}{2024}]{huang2024visioncones}
\begin{botherref}
\oauthor{\bsnm{Huang}, \binits{Z.-F.}},
\oauthor{\bsnm{Vrugt}, \binits{M.t.}},
\oauthor{\bsnm{Wittkowski}, \binits{R.}},
\oauthor{\bsnm{L{\"o}wen}, \binits{H.}}:
Active pattern formation emergent from single-species nonreciprocity.
arXiv preprint arXiv:2404.10093
(2024)
\end{botherref}
\endbibitem

\bibitem[\protect\citeauthoryear{Glauber}{1963}]{glauber1963time}
\begin{barticle}
\bauthor{\bsnm{Glauber}, \binits{R.J.}}:
\batitle{Time-dependent statistics of the ising model}.
\bjtitle{J. Math. Phys.}
\bvolume{4}(\bissue{2}),
\bfpage{294}--\blpage{307}
(\byear{1963})
\end{barticle}
\endbibitem

\bibitem[\protect\citeauthoryear{Bray}{2002}]{bray2002theory}
\begin{barticle}
\bauthor{\bsnm{Bray}, \binits{A.J.}}:
\batitle{Theory of phase-ordering kinetics}.
\bjtitle{Advances in Physics}
\bvolume{51}(\bissue{2}),
\bfpage{481}--\blpage{587}
(\byear{2002})
\end{barticle}
\endbibitem

\bibitem[\protect\citeauthoryear{Cates}{2019}]{cates2019active}
\begin{botherref}
\oauthor{\bsnm{Cates}, \binits{M.E.}}:
Active field theories.
arXiv preprint arXiv:1904.01330
(2019)
\end{botherref}
\endbibitem

\bibitem[\protect\citeauthoryear{Toner and Tu}{1998}]{tonertu1998}
\begin{barticle}
\bauthor{\bsnm{Toner}, \binits{J.}},
\bauthor{\bsnm{Tu}, \binits{Y.}}:
\batitle{Flocks, herds, and schools: A quantitative theory of flocking}.
\bjtitle{Phys. Rev. E}
\bvolume{58},
\bfpage{4828}--\blpage{4858}
(\byear{1998})
\doiurl{10.1103/PhysRevE.58.4828}
\end{barticle}
\endbibitem

\bibitem[\protect\citeauthoryear{Fisher and Barber}{1972}]{fisher_barber_72}
\begin{barticle}
\bauthor{\bsnm{Fisher}, \binits{M.E.}},
\bauthor{\bsnm{Barber}, \binits{M.N.}}:
\batitle{Scaling theory for finite-size effects in the critical region}.
\bjtitle{Phys. Rev. Lett.}
\bvolume{28},
\bfpage{1516}--\blpage{1519}
(\byear{1972})
\doiurl{10.1103/PhysRevLett.28.1516}
\end{barticle}
\endbibitem

\bibitem[\protect\citeauthoryear{Cardy}{1988}]{cardy2012finite}
\begin{botherref}
\oauthor{\bsnm{Cardy}, \binits{J.}}:
Finite-size scaling.
Elsevier
(1988)
\end{botherref}
\endbibitem

\bibitem[\protect\citeauthoryear{Janssen et~al.}{1989}]{janssen1989new}
\begin{barticle}
\bauthor{\bsnm{Janssen}, \binits{H.}},
\bauthor{\bsnm{Schaub}, \binits{B.}},
\bauthor{\bsnm{Schmittmann}, \binits{B.}}:
\batitle{New universal short-time scaling behaviour of critical relaxation
  processes}.
\bjtitle{Zeitschrift f{\"u}r Physik B Condensed Matter}
\bvolume{73},
\bfpage{539}--\blpage{549}
(\byear{1989})
\end{barticle}
\endbibitem

\bibitem[\protect\citeauthoryear{Calabrese
  et~al.}{2006}]{calabrese2006critical}
\begin{barticle}
\bauthor{\bsnm{Calabrese}, \binits{P.}},
\bauthor{\bsnm{Gambassi}, \binits{A.}},
\bauthor{\bsnm{Krzakala}, \binits{F.}}:
\batitle{Critical ageing of ising ferromagnets relaxing from an ordered state}.
\bjtitle{Journal of Statistical Mechanics: Theory and Experiment}
\bvolume{2006}(\bissue{06}),
\bfpage{06016}
(\byear{2006})
\end{barticle}
\endbibitem

\bibitem[\protect\citeauthoryear{Albano et~al.}{2011}]{albano2011study}
\begin{barticle}
\bauthor{\bsnm{Albano}, \binits{E.}},
\bauthor{\bsnm{Bab}, \binits{M.}},
\bauthor{\bsnm{Baglietto}, \binits{G.}},
\bauthor{\bsnm{Borzi}, \binits{R.}},
\bauthor{\bsnm{Grigera}, \binits{T.}},
\bauthor{\bsnm{Loscar}, \binits{E.}},
\bauthor{\bsnm{Rodriguez}, \binits{D.}},
\bauthor{\bsnm{Puzzo}, \binits{M.R.}},
\bauthor{\bsnm{Saracco}, \binits{G.}}:
\batitle{Study of phase transitions from short-time non-equilibrium behaviour}.
\bjtitle{Reports on Progress in Physics}
\bvolume{74}(\bissue{2}),
\bfpage{026501}
(\byear{2011})
\end{barticle}
\endbibitem

\bibitem[\protect\citeauthoryear{Tucci et~al.}{2024}]{tucci2024nonreciprocal}
\begin{botherref}
\oauthor{\bsnm{Tucci}, \binits{G.}},
\oauthor{\bsnm{Golestanian}, \binits{R.}},
\oauthor{\bsnm{Saha}, \binits{S.}}:
Nonreciprocal collective dynamics in a mixture of phoretic janus colloids.
New Journal of Physics
(2024)
\end{botherref}
\endbibitem

\bibitem[\protect\citeauthoryear{Saha and
  Golestanian}{2022}]{saha2022effervescent}
\begin{botherref}
\oauthor{\bsnm{Saha}, \binits{S.}},
\oauthor{\bsnm{Golestanian}, \binits{R.}}:
Effervescent waves in a binary mixture with non-reciprocal couplings.
arXiv preprint arXiv:2208.14985
(2022)
\end{botherref}
\endbibitem

\end{thebibliography}

\end{document}